\documentclass[traditabstract]{aa}
\usepackage{graphicx}
\usepackage{natbib}
\usepackage{amssymb}
\usepackage{amsfonts}
\usepackage{amsbsy}
\usepackage{amsmath}
\usepackage[]{booktabs}

\bibpunct{(}{)}{;}{a}{}{,}

\newcommand{\PFA}{P_{\rm FA}}
\newcommand{\PD}{P_{\rm D}}

\newcommand{\oneb}{\boldsymbol{1}}
\newcommand{\ab}{\boldsymbol{a}}

\newcommand{\gb}{\boldsymbol{g}}
\newcommand{\Cb}{\boldsymbol{C}}

\newcommand{\Db}{\boldsymbol{D}}
\newcommand{\Hb}{\boldsymbol{H}}

\newcommand{\nb}{\boldsymbol{n}}

\newcommand{\ssb}{\boldsymbol{s}}

\newcommand{\xb}{\boldsymbol{x}}

\newcommand{\ub}{\boldsymbol{u}}

\newcommand{\wb}{\boldsymbol{w}}

\newcommand{\Chb}{\boldsymbol{\widehat C}}
\newcommand{\xhb}{\boldsymbol{\widehat x}}
\newcommand{\shb}{\boldsymbol{\widehat s}}

\newcommand{\Hc}{\mathcal{H}}

\newcommand{\Smatb}{\boldsymbol{{\mathcal S}}}
\newcommand{\Xmatb}{\boldsymbol{{\mathcal X}}}
\newcommand{\Ymatb}{\boldsymbol{{\mathcal Y}}}
\newcommand{\Bmatb}{\boldsymbol{{\mathcal B}}}

\newcommand{\Ematb}{\boldsymbol{{\mathcal E}}}

\newcommand{\Fmatb}{\boldsymbol{{\mathcal F}}}
\newcommand{\Gmatb}{\boldsymbol{{\mathcal G}}}
\newcommand{\Nmatb}{\boldsymbol{{\mathcal N}}}

\begin{document}

   \title{Detection of new point-sources in WMAP Cosmic Microwave Background (CMB) maps at high Galactic latitude}
   \subtitle{A new technique to extract point-sources from CMB maps}

   \author{Elsa Patr\'icia R. G. Ramos\inst{1,2,4}
   \and Roberto Vio\inst{3}
    \and
           Paola Andreani\inst{4,5}
          }
   \institute{Centro de Astrof\'isica, Universidade do Porto, Rua das Estrelas, 4150-762 Porto, Portugal\\
             \email{eramos@astro.up.pt}
        \and
	     Departamento de F\'isica e Astronomia da Faculdade de Ci\^encias da Universidade do Porto, Rua do Campo Alegre, 687, 4169-007 Porto, Portugal
        \and
              Chip Computers Consulting s.r.l., Viale Don L.~Sturzo 82,
              S.Liberale di Marcon, 30020 Venice, Italy\\
              \email{robertovio@tin.it},
         \and
		  ESO, Karl Schwarzschild strasse 2, 85748 Garching, Germany
         \and
                  INAF-Osservatorio Astronomico di Trieste, via Tiepolo 11, 34143 Trieste, Italy\\              
		  \email{pandrean@eso.org}
             }

\date{Received .............; accepted ................}

\abstract{
In experimental microwave maps, point-sources can strongly affect the estimation of the power-spectrum and/or the test of Gaussianity 
of the Cosmic Microwave Background (CMB) component. As a consequence, their removal from the sky maps represents a critical step in the 
analysis of the CMB data. Before removing a source, however, it is necessary to detect it and source extraction consists of
a delicate preliminary operation.
In the literature, various techniques have been presented to detect point-sources in the sky maps. The most sophisticated ones
exploit the multi-frequency nature of the observations that is typical of the CMB experiments. These techniques have ``optimal'' theoretical 
properties and, at least in principle, are capable
of remarkable performances. Actually, they are rather difficult to use and this deteriorates the quality of the obtainable results. 
In this paper, we present a new 
technique, the {\it weighted matched filter} (WMF), that is quite simple to use and hence more robust in practical applications. Such technique shows particular efficiency in the detection of sources whose spectra have a slope different from zero. We apply this method to three Southern Hemisphere sky regions -- each with an area of 400 deg$^2$ -- of the seven years Wilkinson Microwave Anisotropy Probe (WMAP) maps and compare the resulting sources with those of the two 
seven-year WMAP point-sources
catalogues. In these selected regions we find seven additional sources not previously listed in WMAP catalogues and discuss their most likely identification and spectral properties.}

\titlerunning{Point-Source Detection}
\authorrunning{E. P. Ramos, R. Vio, \& P. Andreani}
\maketitle

\section{Introduction}
The detection of point-sources embedded in a noise background is a critical issue
in the analysis of the experimental Cosmic Microwave Background (CMB) maps. The estimation of the power-spectrum of the CMB component 
and the test of its possible 
non-Gaussian nature need ``a priori'' detection and removal of these sources. In particular the former operation is rather delicate due to the usually present diffuse background of astrophysical nature and/or the inevitable instrumental noise.
With the increasing high sensitivities of instruments to detect the CMB signal, the astrophysical foregrounds have become the major source of contamination with respect to the instrumental noise. In CMB experiments, the foreground signals at high galactic latitudes come mainly from the emission of extragalactic point-sources.
Given its importance, this subject has been extensively considered in literature
\citep[see e.g.][and references therein]{her08a, car09}. Among the various proposed techniques, the multi-frequency approaches appear to be the most promising 
ones. A good example is the {\it multi-frequency matched filter} (MMF), a well known technique in the community of the ``{\it digital signal processing}''
\citep[e.g. see][]{kay98}, that has been recently proposed by \citet{her02} and \citet{lan10}. Although in principle such technique has ``{\it optimal}''
properties, its use is rather difficult limiting the actual performance in real experimental scenarios. In addition, most of the detection methods available in literature have been developed
in the context of full-sky observations.
These experiments undoubtedly represent an important tool to better understand the physical properties of CMB. However, they suffer the drawback of the Galactic contamination
that, in spite of the optimism expressed in many papers, it could not been yet completely removed.
In this work we present a new technique, that we call {\it weighted matched filter}
(WMF), tailored to the detection of point-sources in regions where the Galactic contamination can be considered negligible. This technique takes also into account the instrumental noise and can be applied to small sky regions.

Observational radio data such as those provided by the Wilkinson Microwave Anisotropy Probe (WMAP) satellite should be a test to the robustness of the WMF to detect extragalactic point-sources. Since this technique can be used to small sky patches, we plan in the future to exploit the high sensitivity and angular resolution of the Atacama Large Millimeter/submillimeter Array (ALMA) and to test its applicability to this facility we plan to apply the WMF to simulated ALMA data (Ramos et al., in preparation).

The paper is divided as follows: in Sec.~\ref{sec:CMB} the mathematics of the WMF is described. Associated numerical experiments are shown in Sec.~\ref{sec:numerical}. The application of the WMF
to the WMAP observational data and the identification of the new discovered sources are presented in Sec.~\ref{sec:wmap}. Conclusions and future development of this work are discussed in Sec. \ref{sec:conclusions}. To speed up the reading of the article most of the mathematical technical details are deferred to the Appendix.

\section{Point-source detection at high Galactic latitude} \label{sec:CMB}

In the context of point-source detection, data can be thought as two-dimensional discrete maps $\{ \Xmatb_i \}_{i=1}^M$, each of them containing
$N_p$ pixels, corresponding to $M$ different observing frequencies (channels), with the form
\begin{equation} \label{eq:observed}
\Xmatb_i = \Smatb_i + \Nmatb_i. 
\end{equation} 
Here, $\Smatb_i$ corresponds to the contribution of the point-sources at the $i$th frequency, whereas $\Nmatb_i$ denotes the corresponding
noise component. At high Galactic latitudes, the CMB component is expected to be the dominant one. Hence, $\Nmatb_i$ may be modeled by
\begin{equation} \label{eq:noise}
\Nmatb_i = \Bmatb + \Ematb_i,  
\end{equation}
where $\Bmatb$ is the contribution of the CMB component 
which is the same at all frequencies (in terms of thermodynamic temperature units) and $\Ematb_i$ is the instrumental noise corresponding to the $i$th channel.

The contribution of the point-sources is assumed to have the form
\begin{equation} \label{eq:point}
\Smatb_i = a_i \Gmatb,
\end{equation}
with $a_i$ the amplitude of the source to the $i$th channel
and where all the sources are assumed to have the same profile $\Gmatb$ independently of the 
observing frequency. Although, in general, this will not be true, it is possible to meet this condition 
by convolving the images with an appropriate kernel.
In the following, the components $\{ \Ematb_i \}$ are considered realizations
of a Gaussian, stationary, zero-mean, stochastic process.

The main feature of model~(\ref{eq:observed})-(\ref{eq:noise}) is that the CMB contribution does not change with the frequency. Hence,
it is possible to linearly combine the maps $\{ \Xmatb_i \} $ in a single map $\Ymatb$
in such a way that 
the CMB contribution is zeroed. In particular,
\begin{equation} \label{eq:weighted}
\Ymatb = \sum_{i=1}^M w_k \Xmatb_i,
\end{equation} 
where the weights $\wb$ are chosen in order to fulfill the criteria
\begin{align}
\wb^T \oneb & = 0, \label{eq:constraint1} \\
\wb^T \ab & = 1, \label{eq:constraint2}
\end{align}
with $\wb=[w_1, w_2, \ldots, w_M]^T$, $\ab=[a_1, a_2, \ldots, a_M]^T$, and $\oneb=(1,1,....1)^T$. 
Here, symbol ``${}^T$'' denotes the {\it matrix transpose}. 
The first constraint (\ref{eq:constraint1}) implies that the contribution of $\Bmatb$ in 
$\Ymatb$ is completely removed,
whereas the second one (\ref{eq:constraint2}) provides a normalizing factor. 
The advantage of this procedure is to deal with a map contaminated only by the instrumental noise. 
It is easier to deal with this kind of noise than with $\Bmatb$ since the correlation length of $\Nmatb$ is much shorter
than that of $\Bmatb$. This is particularly useful in situations where only small
patches of sky are available and the auto-covariance function of the noise (a piece of information necessary to any detection method) 
has to be estimated from the data.
Moreover, working with a single map allows to use detection techniques as the classical {\it matched filter} (MF) 
(see Appendix~\ref{sec:efficient})
whose robustness is proved by many years of applications in many different fields of science and engineering \citep{vio02, vio04}.
We call here {\it weighted matched filter} (WMF) the coupling of the weighted combination of the maps with the MF.

In the case of $M=2$, (i.e. two maps are available), the only possible solution is $\wb^T = [1/(a_1-a_2), -1/(a_1-a_2)]$.
However, for $M > 2$ more degrees of freedom are available. This allows the selection of the weights in
such a way that specific conditions are satisfied. In particular, one could wish that the peak 
signal-to-noise ratio of $\Ymatb$, 
\begin{equation}
R(\wb | \ab) = \frac{(\wb^T \ab)^2}{\wb^T \Db \wb},
\end{equation}
is maximized, i.e.  \footnote{We recall that the functions ``$\arg\min F(x)$'' and ``$\arg\max F(x)$''
provide the values of $x$ for which the function $F(x)$ has the smallest and greatest value, respectively.}
\begin{equation} \label{eq:Rp}
\wb = \underset{ \wb }{\arg\max}  R(\wb | \ab).
\end{equation}
Here, $\Db$ is the $M \times M$ cross-covariance matrix of the noise processes whose $(i,j)$th entry $(\Db)_{ij}$ is given by 
\begin{equation} \label{eq:Dij} 
(\Db)_{ij} = \sigma_{ij}^2,
\end{equation}
with $\sigma_{ii}^2$ the variance of $\Ematb_i$ and  $\sigma_{ij}^2$ the covariance between $\Ematb_i$ and $\Ematb_j$.
Because of the constraint (\ref{eq:constraint2}), condition~(\ref{eq:Rp}) can be reformulated as
\begin{equation} \label{eq:R}
\wb = \underset{ \wb }{\arg\min} [{\wb^T \Db \wb}].
\end{equation}

This approach differs 
from that proposed by \citet{che09} which consists in the minimization of the simpler quantity $\wb^T \wb$ 
(i.e. instrumental noise is not taken into account). Moreover,
these authors seem to adopt a numerical approach for such operation (no details are provided respect to this).
Actually,
a simple analytic solution of problem~(\ref{eq:R}), with the constraints~(\ref{eq:constraint1})-(\ref{eq:constraint2}),  
can be obtained by means of the {\it Lagrange multipliers} method, i.e.
\begin{equation} \label{eq:solutionw}
\wb = \frac{\eta \Db^{-1} \ab - \zeta \Db^{-1} \oneb}{\vartheta \eta - \zeta^2},
\end{equation}
where
\begin{align}
\eta & = \oneb^T \Db^{-1} \oneb; \\
\zeta & = \ab^T \Db^{-1} \oneb; \\
\vartheta & = \ab^T \Db^{-1} \ab. 
\end{align}
This result is similar to that obtained by \citet{rem10} and \citet{hur10} however, in a completely different context, in which the problem of interest is 
the separation of CMB and Sunyaev-Zel'dovich effect  
through the {\it internal linear composition} approach. The main difference with the solution~(\ref{eq:solutionw}) is that in \citet{rem10}
and \citet{hur10} the empirical covariance matrix of the observed maps $\Xmatb_i$ is used, instead of the matrix $\Db$ in (\ref{eq:solutionw}).

\section{Numerical experiments} \label{sec:numerical}

To test the performances of the WMF we have carried out some numerical experiments. We consider a scenario
where four different observing frequencies are available. We make the simplifying assumption that all the channels
have the same point-spread function (PSF) which is a two-dimensional circular symmetric Gaussian normalized to have a peak value equal to one and 
with a dispersion set to three pixels. Here the CMB component $\Bmatb$ is simulated on a regular two-dimensional grid containing $(101 \times 101)$ 
pixels with size $3.52' \times 3.52'$. The instrumental noise $\Ematb_i$ is assumed to be a Gaussian white-noise
process with variance equal to one in units of the standard deviation of the CMB signal. 
This scenario mimics that expected for the "Low-Frequency Instrument" mounted on the PLANCK satellite \citep{vio03}. The
amplitudes 
$\{ a_i \}$ of the point-sources are assumed to follow a power-law
\begin{equation} \label{eq:a}
a_i = \left( \frac{\nu_i}{\nu_1} \right)^{\alpha}  a_1
\end{equation} 
where $\nu_i$, $i=1,2,3,4$ are the observing frequencies  
($i = 1 \rightarrow 30$ GHz, $i = 2 \rightarrow 44$ GHz, $i = 3 \rightarrow 70$ GHz, $i=4 \rightarrow 100$ GHz),
$a_1 = 0.5$ (in units of the standard deviation of the CMB signal) and
$\alpha$ is the source spectral index. The value of $a_1$ has been chosen to reproduce an experimental situation characterized by a 
rather low signal-to-noise ratio.

\begin{figure*}
        \resizebox{\hsize}{!}{\includegraphics{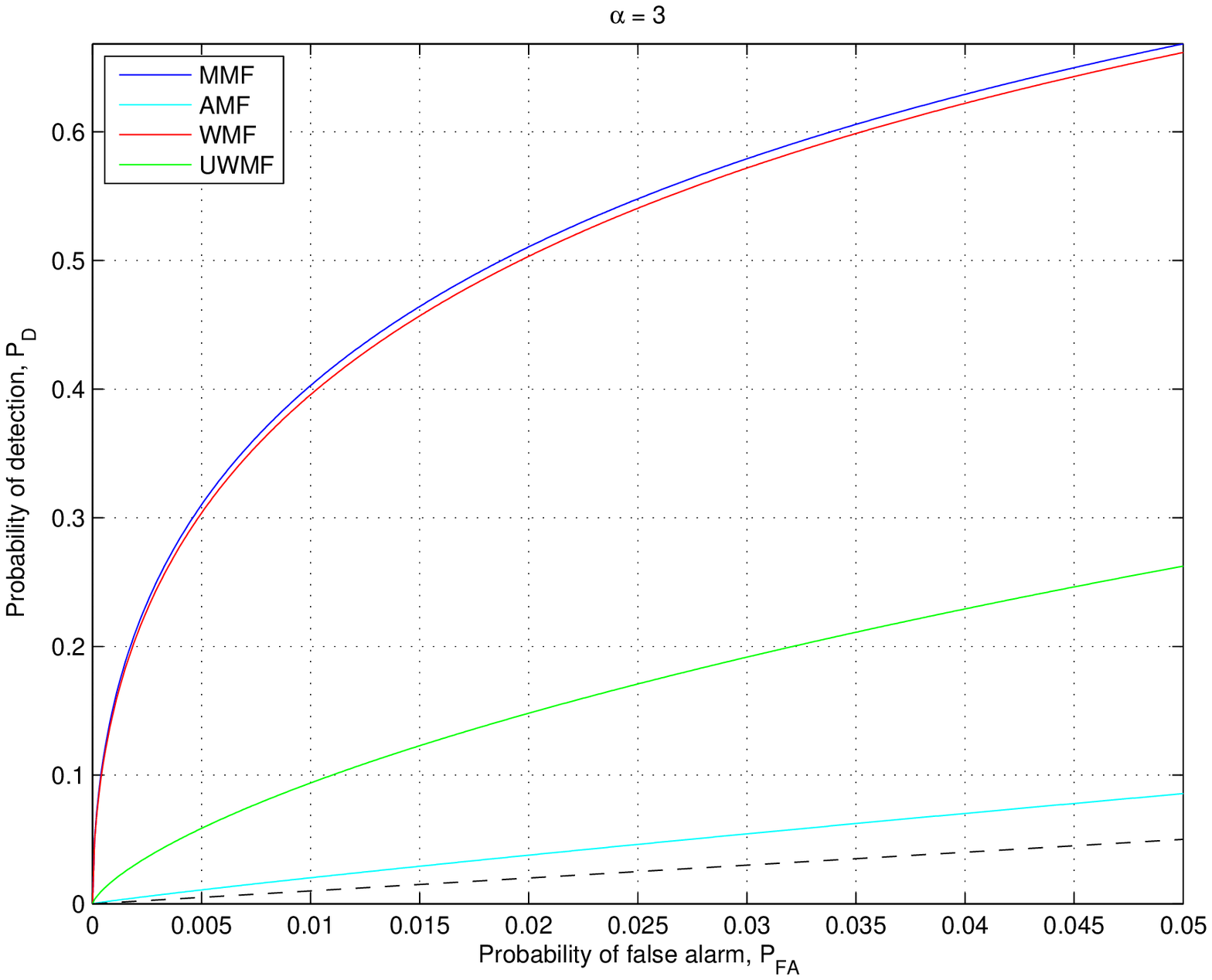}\includegraphics{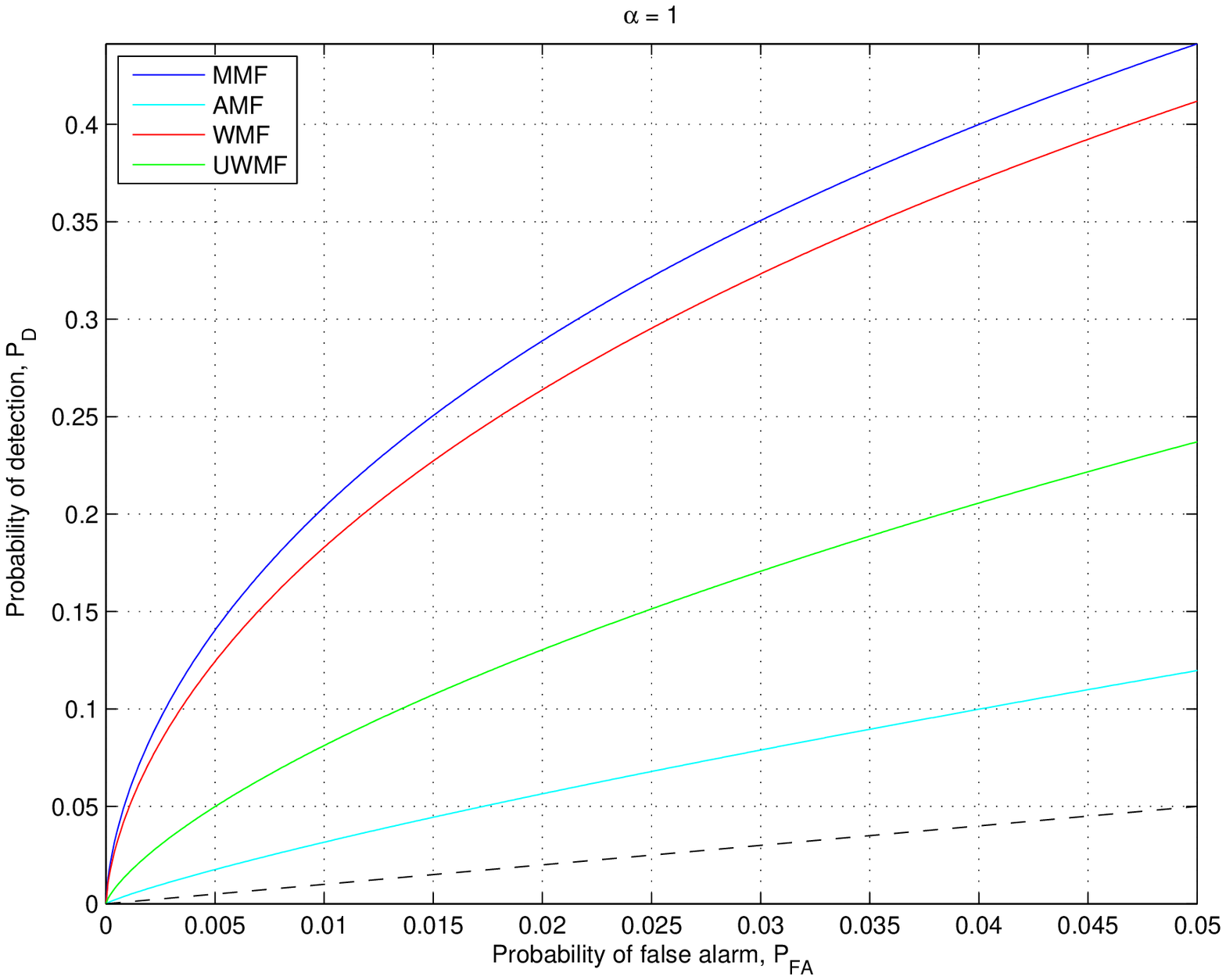}}
        \resizebox{\hsize}{!}{\includegraphics{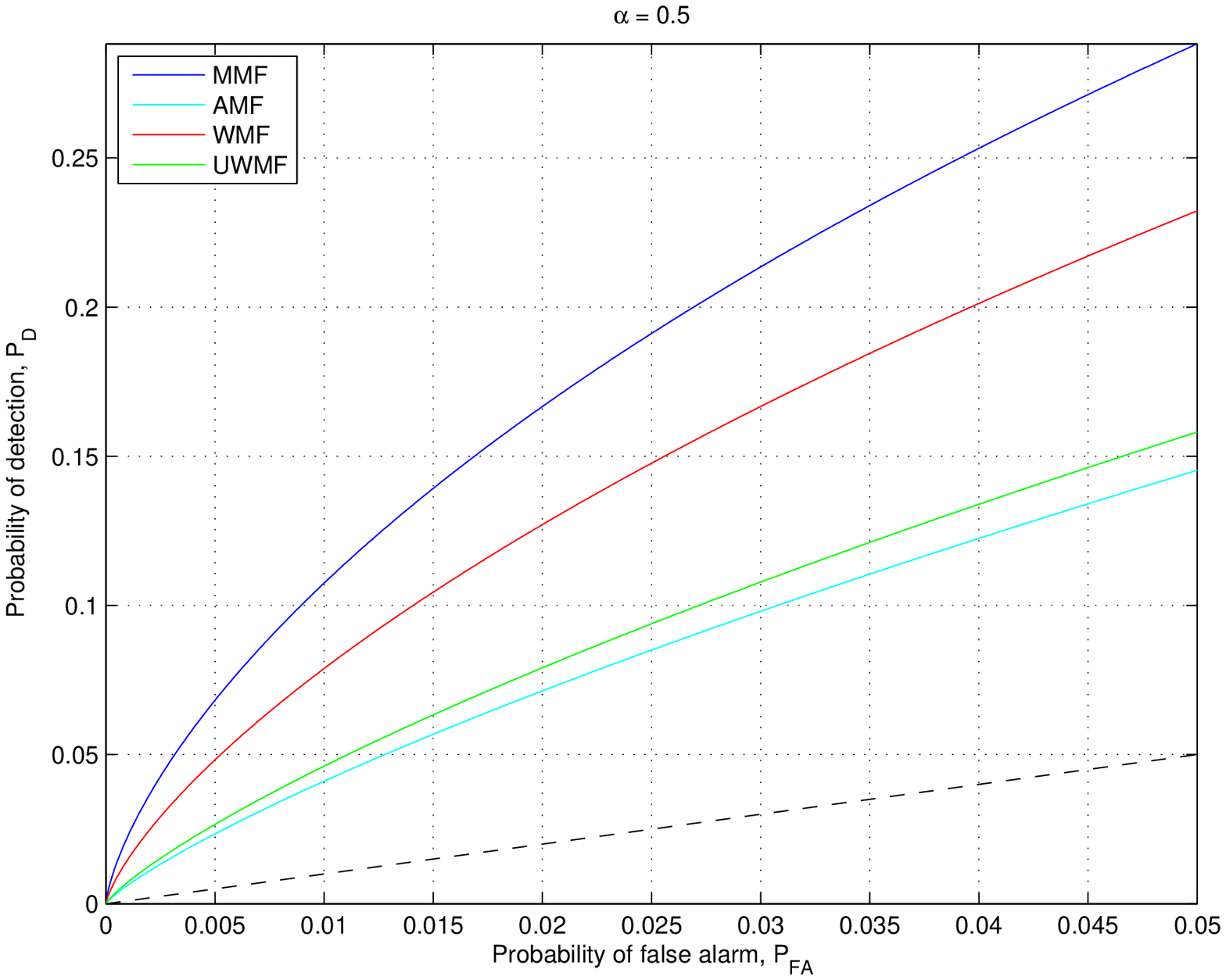}\includegraphics{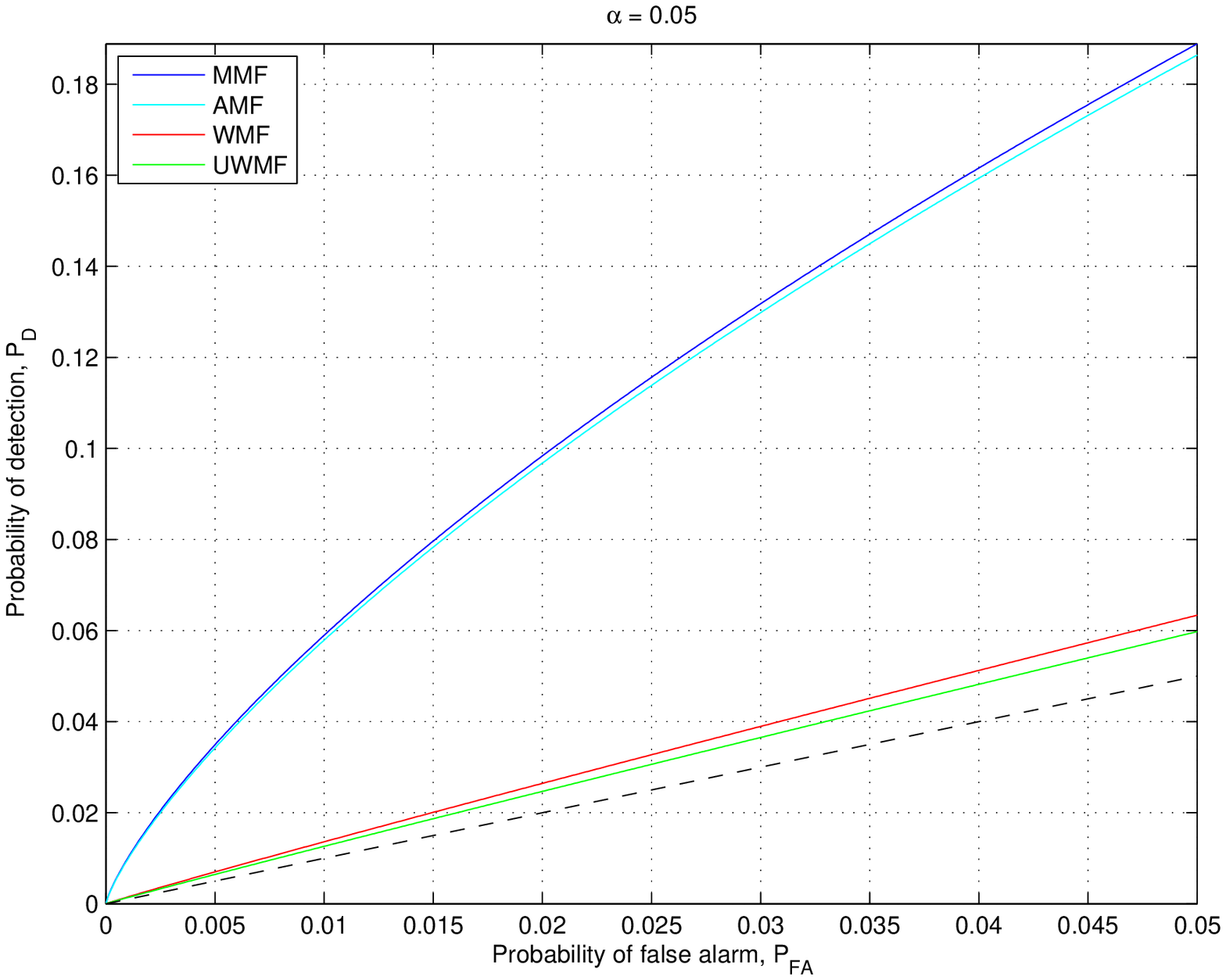}}
        \caption{{\it Probability of detection} $\PD$ against {\it probability of false alarm} $\PFA$ for the detection methods in the numerical 
        experiment described in Sec.~\ref{sec:numerical}. Here, the sources are assumed to have intensity $a_i$ at the $i$-th frequency given by $a_i = (\nu_i / \nu_1)^{\alpha} a_1$, with $a_1=0.5$. Four values for $\alpha$ are considered, $\alpha=3, 1, 0.5, 0.05$.
        The instrumental noise is assumed to be a Gaussian, zero-mean, white-noise process whose standard
        deviation is equal to one in units of the standard deviation of the CMB signal. 
        The results are shown for the different methods: the {\it weighted matched filter} (WMF),  
        the {\it multi-frequency matched filter} (MMF),
        the {\it uniformly weighted matched filter} (UWMF) and the {\it average matched filter} (AMF).
The MMF is used as benchmark since it has the best theoretical detection performance.
        The UWMF shows the worst possible results obtainable with the WMF approach.
        The AMF shows what results are obtainable when the maps are simply averaged. A method is superior to another one when, for a fixed $\PFA$, the
        corresponding $\PD$ is greater (for a given method, the relationship $\PD$ against $\PFA$ should always be well above a $45^{\circ}$ 
        straight line, the dashed line in the figure). With increasing of $\alpha$, the behavior of MMF and WMF becomes similar. 
For an $\alpha$ close to zero, the MMF and AMF show a very similar performance.}
        \label{fig:test}
\end{figure*}

Figure~\ref{fig:test} shows the ``{\it probability of detection}'', $\PD$, against the ``{\it probability of false alarm}'', $\PFA$ (i.e. the probability
of a {\it false detection}), for the WMF  (see Appendix~\ref{sec:efficient}). 
Four different values of $\alpha$ are considered, i.e. $\alpha=3, 1, 0.5, 0.05$ (negative values of $\alpha$ provide similar results).
For comparison we show also the results obtained with the classical {\it multi-frequency matched filter} (MMF), and those obtained
using the WMF with the weights
$\wb = [\rho, \rho, \ldots, -(M-1) \rho]^T$, $\rho = 1 / \sqrt{(1-M) + (1-M)^2}$. This last method, that we name 
{\it uniformly weighted matched filter} (UWMF), corresponds to a situation 
where only one signal is used to eliminate the component $\Bmatb$, whereas the others are given
an identical weight. Quantity $\rho$ is fixed in such a way that $\wb^T \oneb = 0$ and $\wb^T \wb = 1$.
In this experiment MMF and UWMF are used as, respectively, upper and lower limit for the results obtainable by WMF.
This is because, under the conditions we are working with, no detection method can outperform MMF (see Appendix~\ref{sec:efficient}).
On the other side, in general, it is not expected that UWMF achieves good performances, since the weights are computed without considering
the noise level in the map as well the characteristics of the source spectra.
In this kind of diagram, a method is superior to another one when, for a fixed $\PFA$, the
corresponding $\PD$ is greater. More specifically, the relationship $\PD$ against $\PFA$ should always be well above a $45^{\circ}$ 
straight line (the dashed line in the figure) since this corresponds to a detection performance identical to that of flipping a coin, 
ignoring all the data. 

From Fig.~\ref{fig:test} it is evident that, when $\alpha>1$, the WMF and MMF have very similar performances. When $0.5 \leq \alpha \leq 1$, the behavior of these both filters is still reasonably similar. In the case of
$\alpha \approx 0$ (i.e. for sources with flat spectrum) the performance of the WMF becomes close to that of the UWMF. 
This happens because for $\alpha$ close to zero, not only the CMB has the same contribution at the various frequencies but also the intensity of the sources is constant. In this case,
a simple alternative is to average the maps and then apply the classic MF to the resulting map. We call this method
{\it average matched filter} (AMF). A point to stress is that, similarly to the MMF,
also the AMF requires the knowledge
of the auto-covariance matrix of the CMB. This last method, however, has the advantage over the MMF because only one map has to
be handled. For comparison, in Fig.~\ref{fig:test}, the performance of the AMF is also shown. 

\begin{figure*}
        \resizebox{\hsize}{!}{\includegraphics{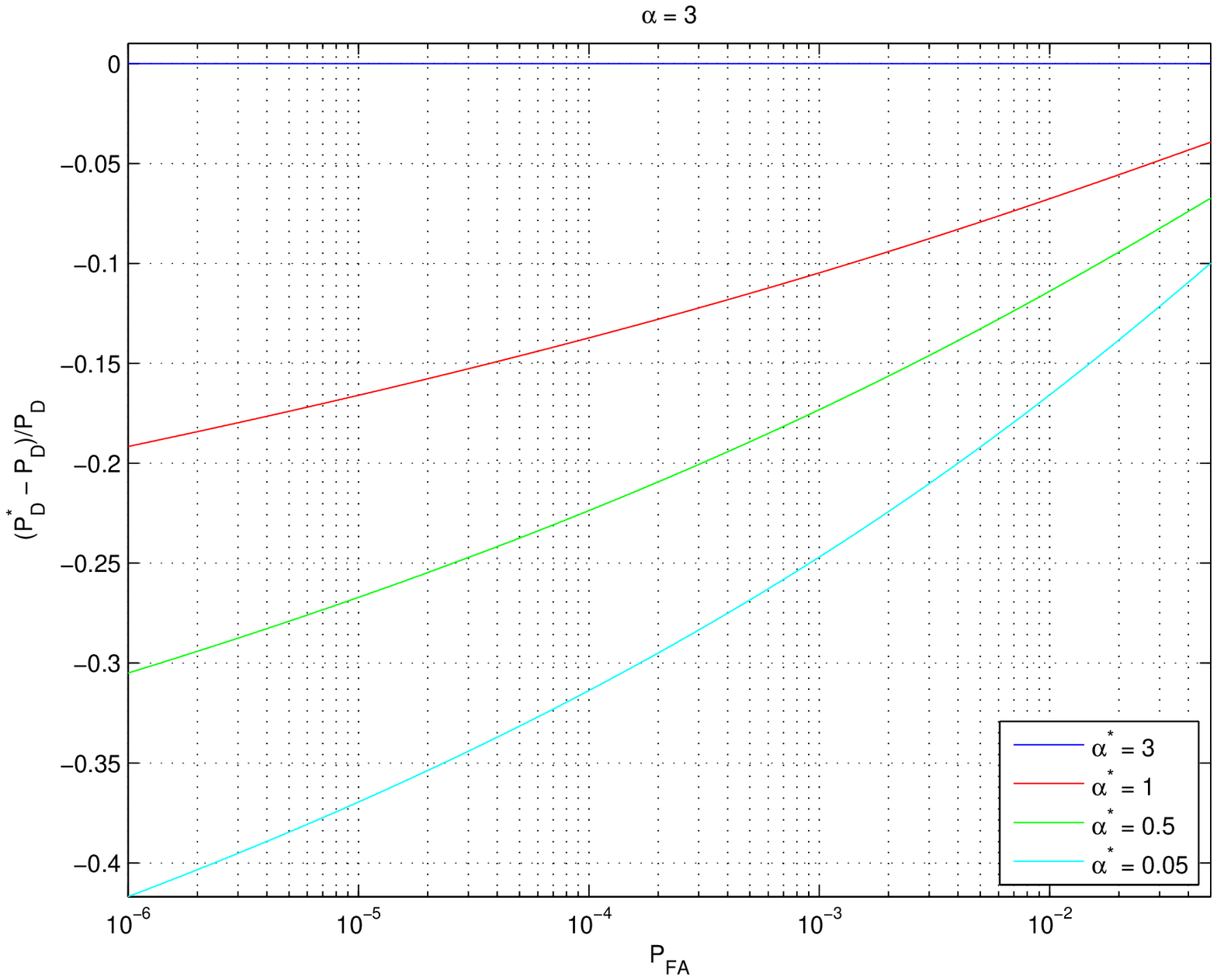}\includegraphics{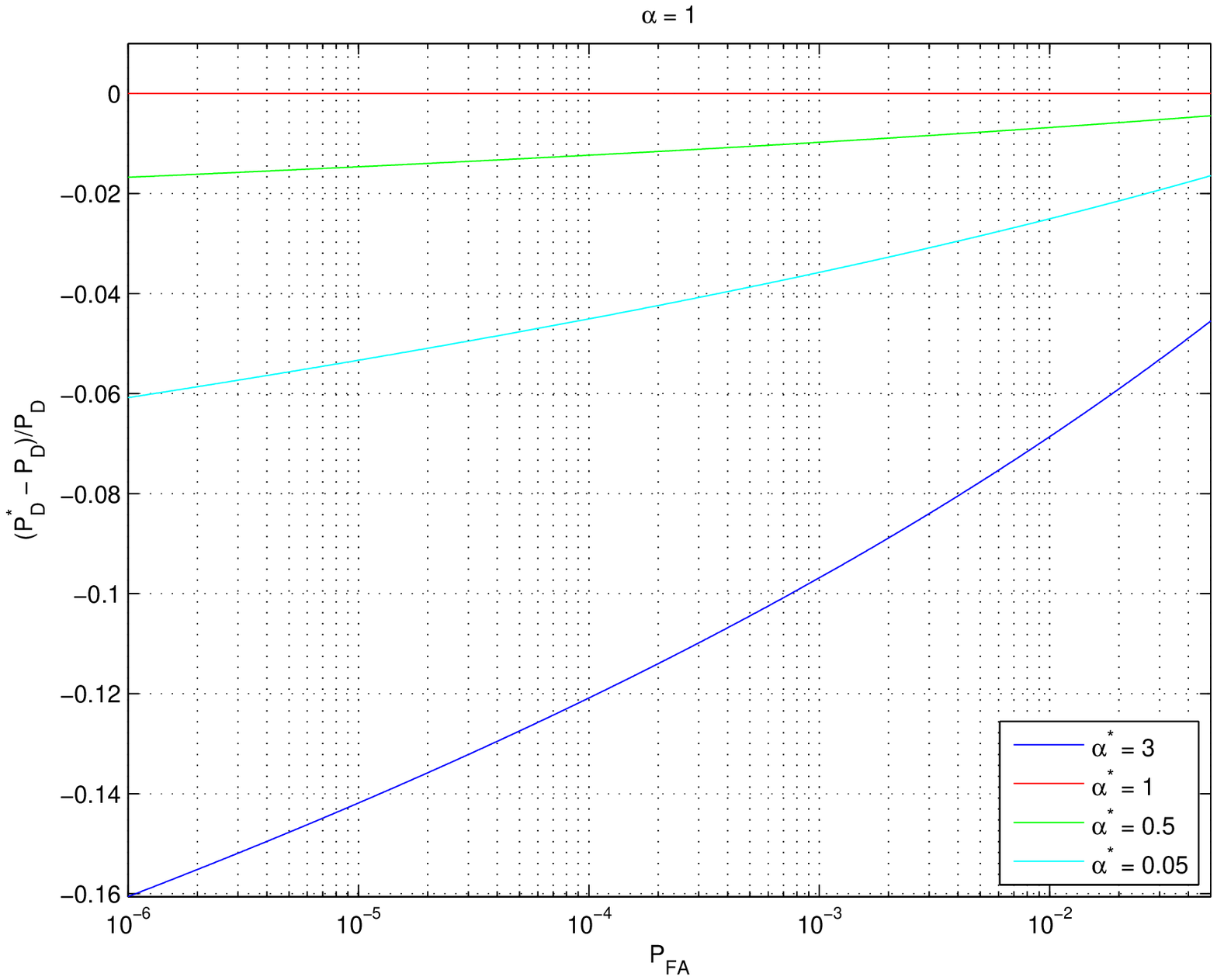}}
        \resizebox{\hsize}{!}{\includegraphics{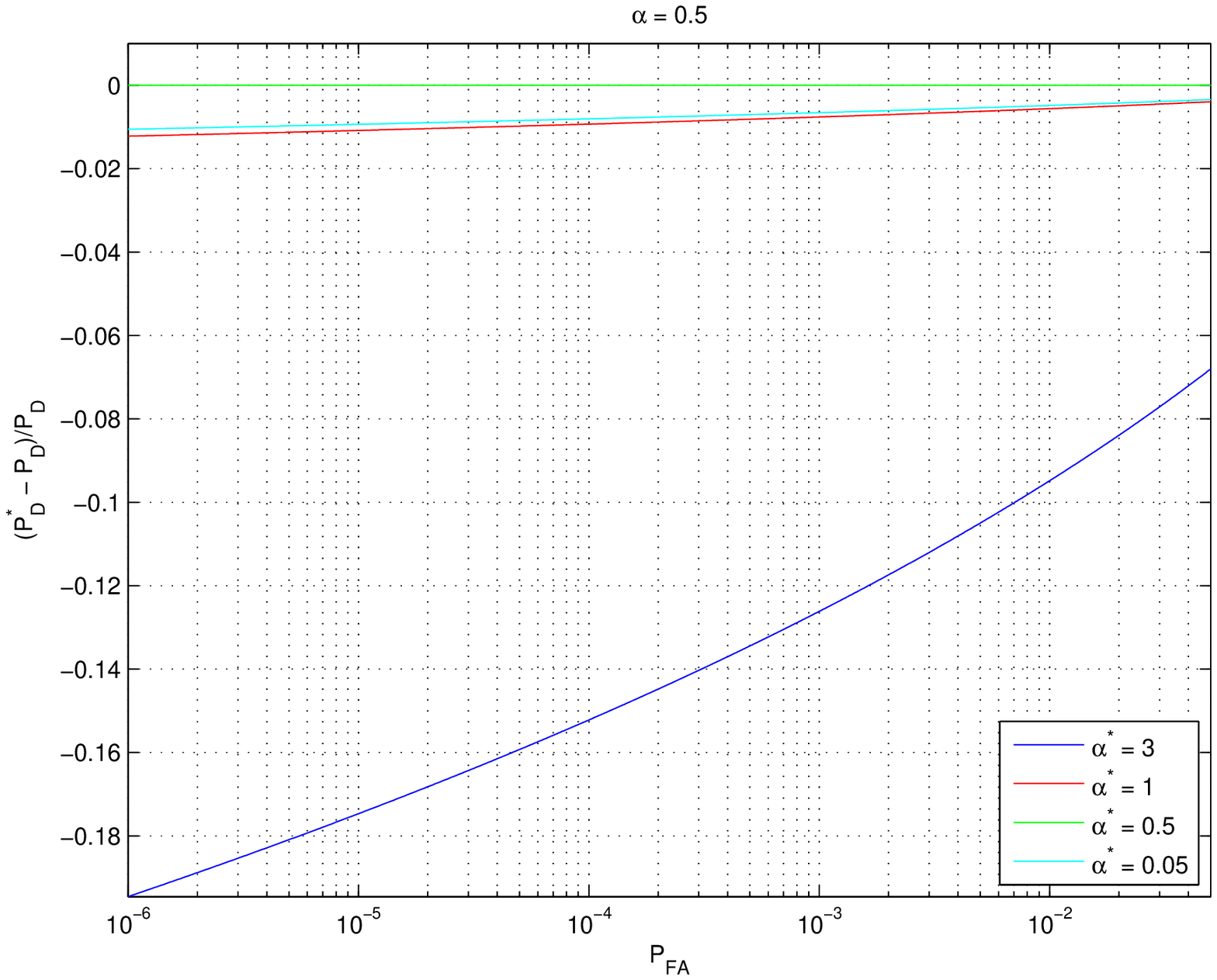}\includegraphics{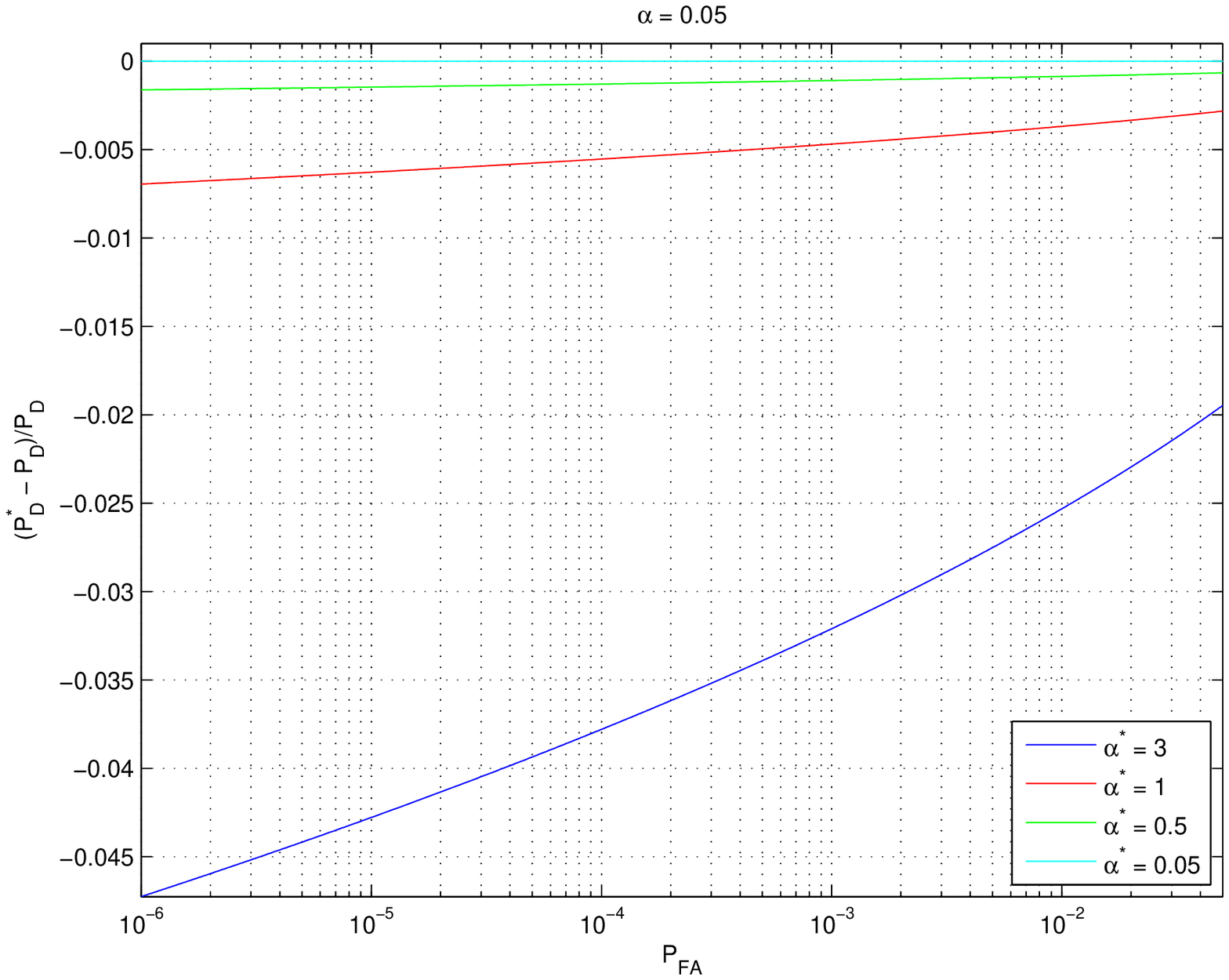}}
        \caption{Relative decrease of the {\it probability of detection}, $(\PD^* - \PD)/\PD$, against $\PFA$ when the WMF is applied to a source
whose true {\it spectral index} $\alpha$ is erroneously assumed to be $\alpha^*$. Here, $\PD$ and $\PD^*$ are the {\it probability of detection}
when the WMF is applied assuming the true and the wrong {\it spectral index}, respectively. The set of values $[3, 1, 0.5, 0.05]$ is used for both 
$\alpha$ and $\alpha^*$.}
        \label{fig:err}
\end{figure*}

From these considerations, it appears that an effective and the simplest procedure to detect less common sources
(those with spectra different from flat) consists of using the WMF, optimised for different values of $\alpha$. The reliability of such procedure is supported by Fig.~\ref{fig:err}
that shows the relative decrease of the {\it probability of detection}, $(\PD^* - \PD)/\PD$, against $\PFA$ when the WMF is applied to a source
whose true {\it spectral index} $\alpha$ is erroneously assumed to be $\alpha^*$. Here, $\PD$ and $\PD^*$ are the {\it probability of detection}
when the WMF is applied assuming the true and the wrong {\it spectral index}, respectively. The set of values $[3, 1, 0.5, 0.05]$ is used for both 
$\alpha$ and $\alpha^*$. From this figure, it is evident that a remarkable decrease of the {\it probability of detection} is to be expected 
only if $\alpha$ is quite different from $\alpha^*$. 

\section{Application to WMAP data} \label{sec:wmap}
\subsection{WMAP maps}
The Wilkinson Microwave Anisotropy Probe (WMAP) satellite (Bennett et al. 2003a) was designed to produce microwave full-sky maps of 
the CMB radiation. With the aim of separate the CMB signal from foreground components, the maps were 
obtained at five different frequency bands, respectively centered at 23GHz (K band), 33 GHz (Ka band), 41 GHz (Q band), 61 GHz (V band) 
and 94 GHz (W band). Although the limited angular resolution of WMAP, with a full width at half maximum (FWHM) of about 13 arcminutes (W band), 
it is presently the only one that can offer a millimeter wavelengths all-sky survey, providing an unique tool for the study of radio sources. 

The seven-year full sky temperature and polarization maps per frequency band, namely the Stokes I, Q and U parameters,
are available from the LAMBDA 
website\footnote{http://lambda.gsfc.nasa.gov/}. The maps of the five frequency bands have different 
resolutions, from roughly 0.21$^\circ$ (W band) to about 0.82$^\circ$ (K band). With the goal of testing the proposal WMF method to identify 
radio point-sources we use the Stokes I (temperature) co-added maps (combination of the individual differencing assemblies of a single frequency 
band) to a common 1$^\circ$ FWHM Gaussian beam, from which the CMB dipole has been removed (for more details see Jarosik et al. 2010). 
These maps were generated as a nested HEALPix\footnote{http://healpix.jpl.nasa.gov/} sky projection (G\'orski et al., 2005) with a resolution 
of Nside=512 (corresponding to the label WMAP resolution of Res 9).

\subsection{Selection of the sky regions}

\begin{figure*}
        \resizebox{\hsize}{!}{\includegraphics{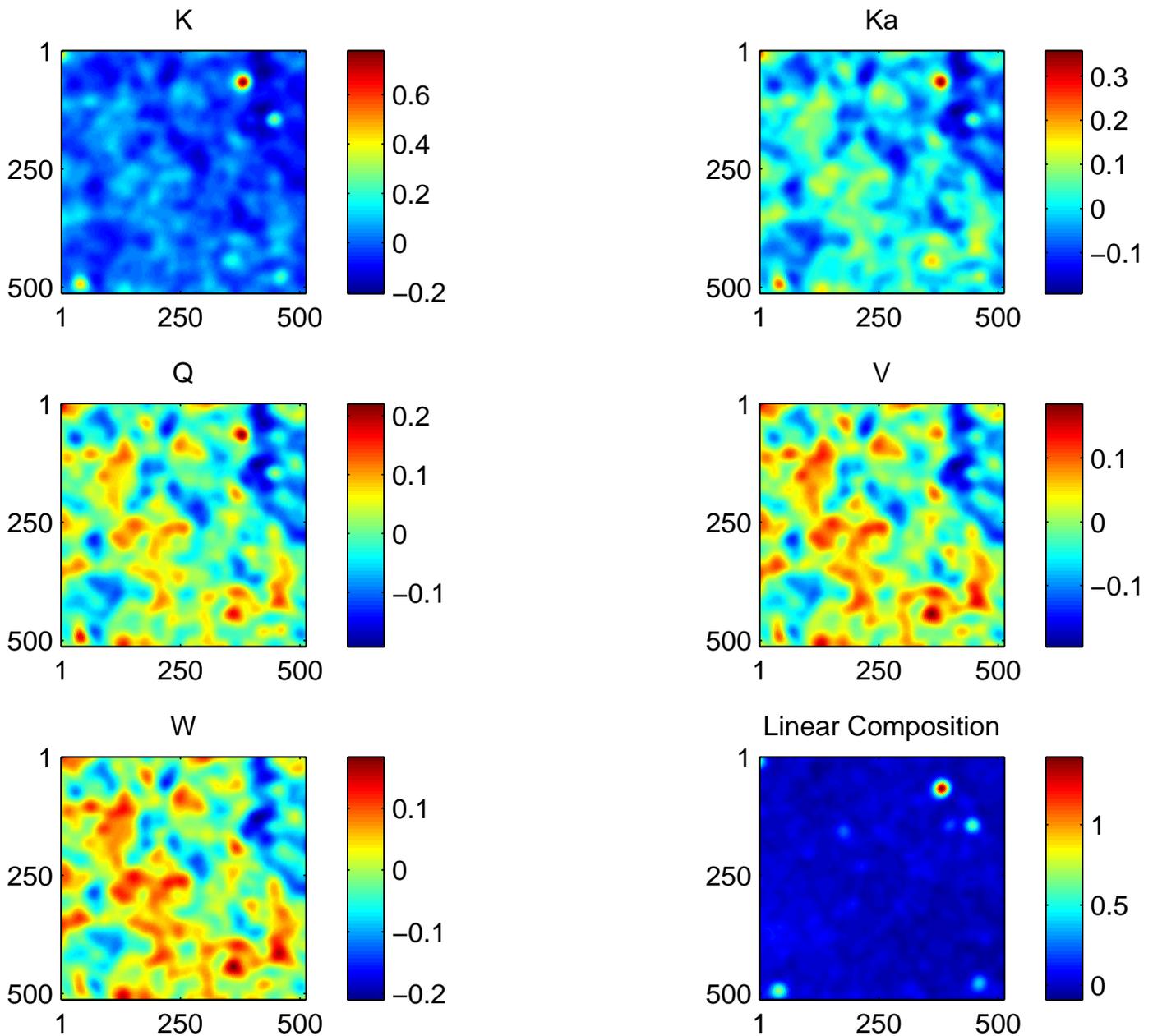}}
        \caption{Squared WMAP maps per frequency band for the first sky region, centered at (l,b)=(258.18$^\circ$,-46.33$^\circ$) with an area of 20$^\circ$$\times$20$^\circ$. The map at the lower right corner corresponds to the linear composition map obtained with the WMF.}
        \label{fig:all_region1}
\end{figure*}

\begin{figure*}
        \resizebox{\hsize}{!}{\includegraphics{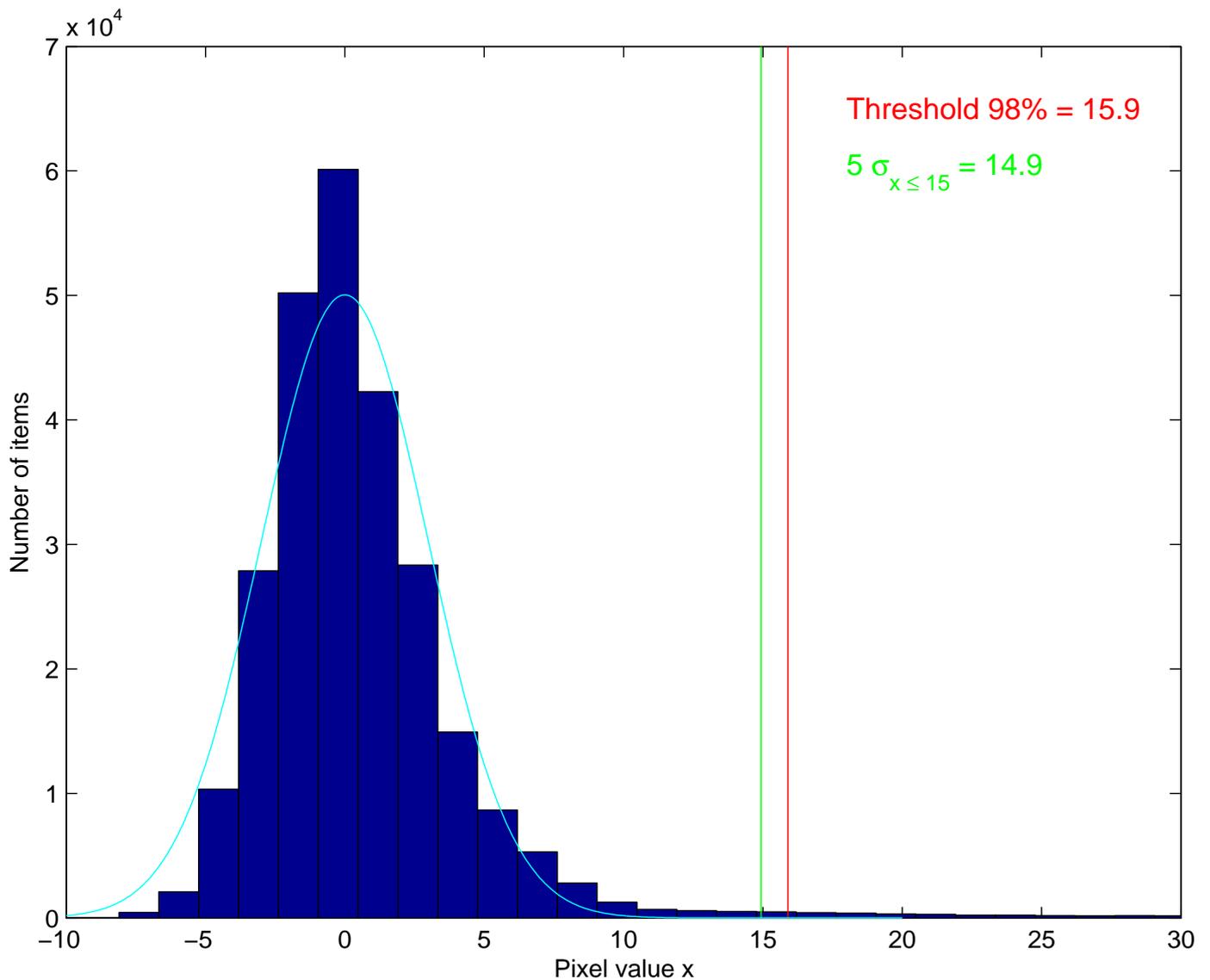}}
        \caption{Empirical probability density function (EPDF) of the values of the pixels for the linear composition map of the first sky region after the application of the matched filter. The EPDF is different from zero in the range $[-11, ~132]$. Here, only the range $[-10,~30]$ is shown since outside such interval the EPDF is close to zero. The vertical red and green lines show the detection threshold based on the $98$th percentile and the $5 \sigma_{15}$ level, respectively. Here, $\sigma_{15}$ is the standard deviation of the pixels with values less or equal to $15$. The cyan line provides the Gaussian probability density function with zero mean and standard deviation given by $\sigma_{15}$. }
        \label{fig:fig_thr}
\end{figure*}

The drawback to use the CMB experiments for cosmological studies is the foreground contamination from the Galaxy and extragalactic
sources. In particular, the extragalactic point-sources contaminate the CMB maps at frequencies below 60 GHz and at high frequencies the
statistical properties are still barely known. At high Galactic latitudes ($|$b$|{>}$15$^\circ$) and for frequencies between 30 and 150
GHz, the CMB signal dominates the Galactic one (e.g. Bennett et al. 2003b, Tegmark et al. 2000). Therefore, a strategic way to obtain
more accurate cosmological information is to observe at high Galactic latitudes where the foreground contamination is expected to be
lower. We selected three particular sky regions, each with an area of 20$^\circ$$\times$20$^\circ$
centered at galactic longitude (l) and latitude (b) of (l,b)=(258.18$^\circ$,-46.33$^\circ$), (l,b)=(252.07$^\circ$,-38.78$^\circ$) and
(l,b)=(272.48$^\circ$,-54.63$^\circ$) denoted here by first, second and third sky region, respectively. The first coordinates were chosen because this region is of particular interest to the EBEx experiment \citep{Reichborn-Kjennerud}, towards which
we plan follow-up observations with the Atacama Large Millimetre/submillimetre Array (ALMA). The centres of the other two regions are two point-sources identified in the first one. We extracted these three sky regions from the smoothed full-sky maps per frequency band and project them in squared maps of ($512 \times 512$) pixels
using the HEALPix software. In Fig.~\ref{fig:all_region1} the resulting maps in all the frequency bands for the first region are shown. 
We applied the WMF method to the three regions mentioned above and the linear composition map for the first region can be seen in Fig.~\ref{fig:all_region1}. 
To fix the detection threshold, we consider an approach based on the {\it empirical probability density function} (EPDF) of the values of the pixels of the linear composition maps after the application of the matched filter. This is the typically procedure used for the detection of point-sources in the CMB context. Many authors set such threshold to five times the standard deviation of the pixel values in the map ($5\sigma$ level). This method, however, suffers the impact of the point-sources themselves \citep[e.g. see][]{lea08}. For this reason, we have adopted a different approach that is much less dependent on the amount of the spurious contributions. 
In particular, we claim a detection when the fluxes in the corresponding pixels
have values above a threshold given by the $98\%$ percentile computed over all the pixels in the respective sky region.
The choice of this approach is forced by the difficulties in the computation of the detection threshold starting from an ``a priori'' $\PFA$ 
since we could not used the level of the pixel noise provided by the WMAP team  (defined as $\sigma=\sigma_0/ \sqrt {N_{\rm obs}} $, where $\sigma_0$ and $N_{\rm obs}$ are values taken from the LAMBDA website \footnote{http://lambda.gsfc.nasa.gov/product/map/dr4/}).
The reason is that, before their linear composition, the maps have been manipulated to obtain a common spatial resolution as well as to convert them from spherical to rectangular coordinates (our codes are developed for the case of small patches of sky). Moreover, the noise in the WMAP maps is not spatially uniform. The  EPDF for the first region can be seen in Fig.~\ref{fig:fig_thr}.
It is evident that the bulk of the pixel values is confined in a rather restricted range, say $[-10,~ 15]$ in internal units of our codes with a standard deviation of about $\sigma_{15}=2.98$ units. 
Since the CMB component is not present in the linear composition map, it is reasonably to assume that these values are due only to the noise. The EPDF presents a long tail up to a value of about $132$ units that is due to the presence of the point-sources. The $98\%$ empirical percentile of the EPDF roughly corresponds to the  $5\sigma_{15}$ level. Similar values are found for the other regions.

\subsection{Identification of the WMAP Point-Sources}

The WMAP seven-year Point Source Catalogs contains information on the point-sources in the five frequency bands from 23 to 94 GHz, based on data
from the first 7 years of the WMAP sky survey from 10 Aug 2001 to 9 Aug 2008, inclusive.
The WMAP team has produced two point-source catalogues using different methods for the identification of the sources,
namely the Five-band search technique and the three-band CMB-free technique
\footnote{http://heasarc.gsfc.nasa.gov/W3Browse/all/wmapptsrc.html}. The former catalogue contains 471 point-sources and it is
complete to 2 Jy for regions of the sky away from the Galactic
plane. The latter catalogue was built using the three frequency bands from 41 to 94 GHz. This last method identifies 417 point-sources in a linear combination map for which the weights were obtained such that the CMB contribution was removed and point-sources with flat-spectrum were preserved. 

\begin{figure*}
        \resizebox{\hsize}{!}{\includegraphics{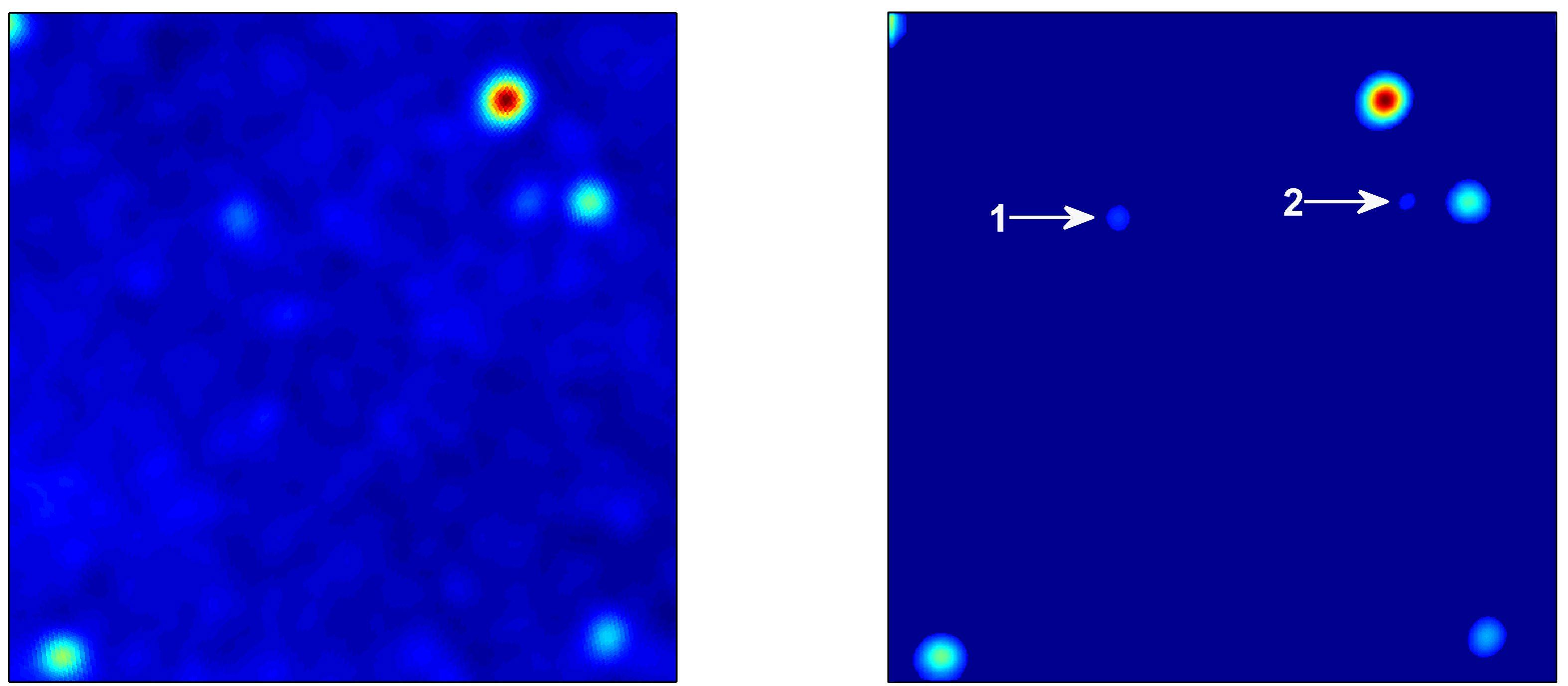}}
        \caption{Left Panel -- Linearly composed image obtained with the WMF for the first sky region centered at (l,b)=(258.18$^\circ$,-46.33$^\circ$); Right Panel -- To enhance the source appearence it is shown the same figure with the pixels with
the smallest values ($98\%$ of the total) zeroed.}
        \label{fig:region1}
        \resizebox{\hsize}{!}{\includegraphics{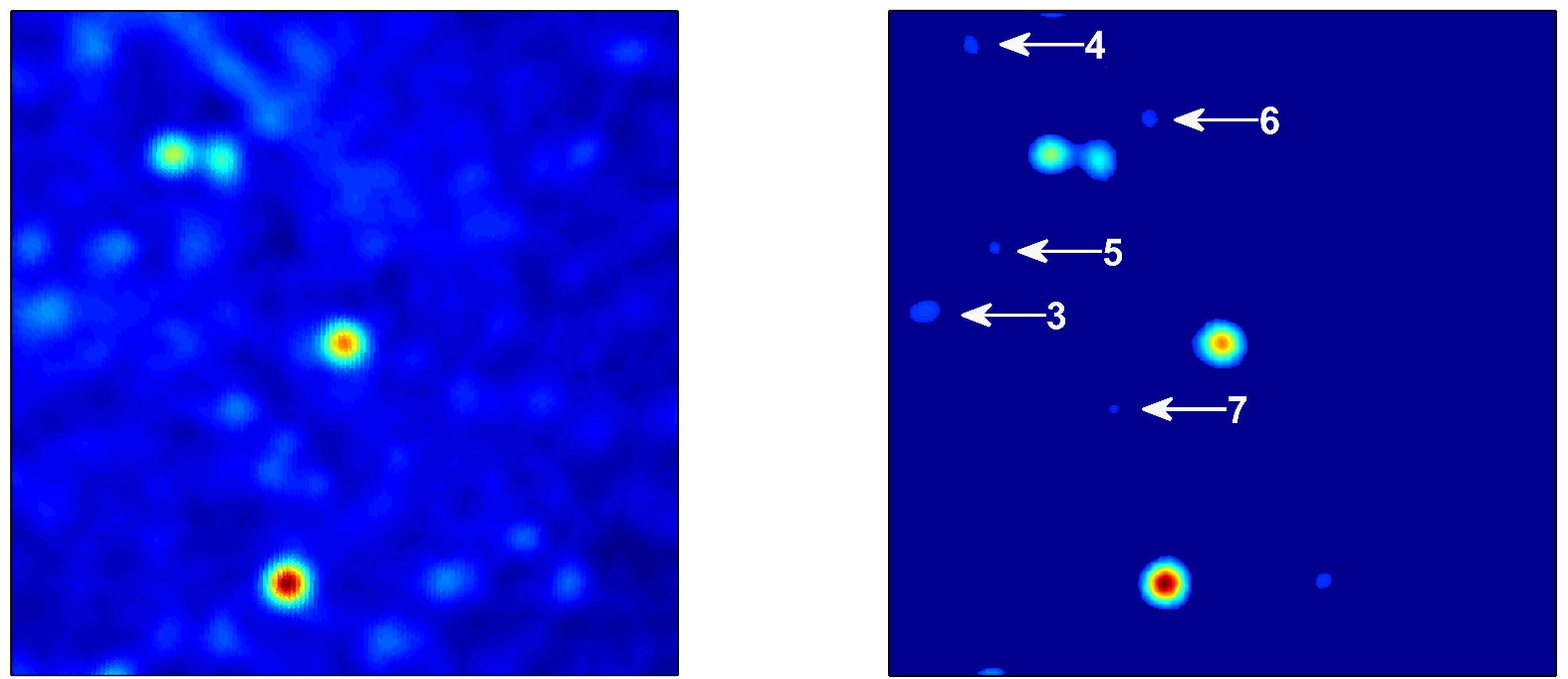}}
        \caption{Same as Fig.~\ref{fig:region1} for the third sky region centered at (l,b)=(272.48$^\circ$,-54.63$^\circ$).}
        \label{fig:region3}
\end{figure*}

In the three sky regions selected for this work all the sources found with the WMF were
cross-checked with those found in both WMAP seven-year catalogues. 
All the sources found by the WMAP team for the selected regions are detected by WMF. However, the WMF allows us to find more
sources that are not listed in none of these catalogues. Those new sources can be seen in the Fig.~\ref{fig:region1} and Fig.~\ref{fig:region3}. We have labeled them as {\it Source 1} and {\it 2} in the first region and {\it 3}, {\it 4}, {\it 5}, {\it 6} and
{\it 7} in the third region. All the sources detected by the WMF in the second
region are listed already in the WMAP catalogues, except two ({\it Source 1} and {\it 2}) that are already identified in the first region.

As our detection method is sensitive to the source spectral index, we investigated what kind of spectral type
radio source populations were detected by WMAP. The Five-band catalogue provides an estimate of
the spectral index ($\alpha$) defined by a power law of the form, F $\propto$ $\nu^{-\alpha}$, where F is the flux density
and $\nu$ is the frequency. We made use of this information to study the distribution of the spectral indices. The
$\alpha$ lies in the range [-2.1,1.3] and three main spectral classes can be identified:
about 81\% of the population have a flat-spectrum (-0.5${\leq}\alpha{\leq}$0.5),
16\% show an inverted-spectrum ($\alpha{<-0.5}$) and the remaining 3\% a steep-spectrum ($\alpha{>}0.5$).

We test different values for $\alpha$ with the WMF and the new sources are detected for spectral indices in the range obtained from the Five-band catalogue, except in the case of assuming an $\alpha$ close to 0. As said in Sec.~\ref{sec:numerical} the WMF is more efficient in finding sources with spectral index different from zero. For strictly flat-spectrum sources ($\alpha\simeq 0$) the method
erases not only the CMB component but also the sources having equal intensity in all the WMAP bands. Therefore, this method is optimized for those sources having a flux dependent on the frequency.  

\subsubsection{Sources fluxes}

To identify the new detected sources we need to have an estimation of the fluxes measured by WMAP. To this aim we used the
original seven-years maps and integrate the flux density at the source location within the WMAP beam, assumed to have a Gaussian
profile with FWHM as given in Hinshaw et al. (2009). Conversion factors from mK to Jy have been derived
by comparing our derived fluxes with those given in the WMAP catalogues. 

\subsubsection{Cross identification of the new discovered WMAP Point-Sources}

To identify possible counterparts we cross-correlated the new WMAP sources with catalogues found in the NED database, AT20G Catalogue
(Murphy et al., 2010) and NEWPS sources (Massardi et al., 2009). We checked all
the sources with a radio counterpart within a selected radius of 12$^\prime$ which corresponds to 3 times the mean position uncertainty of the WMAP satellite (Gold et al., 2010).

\begin{figure*}
	\includegraphics[height=8cm]{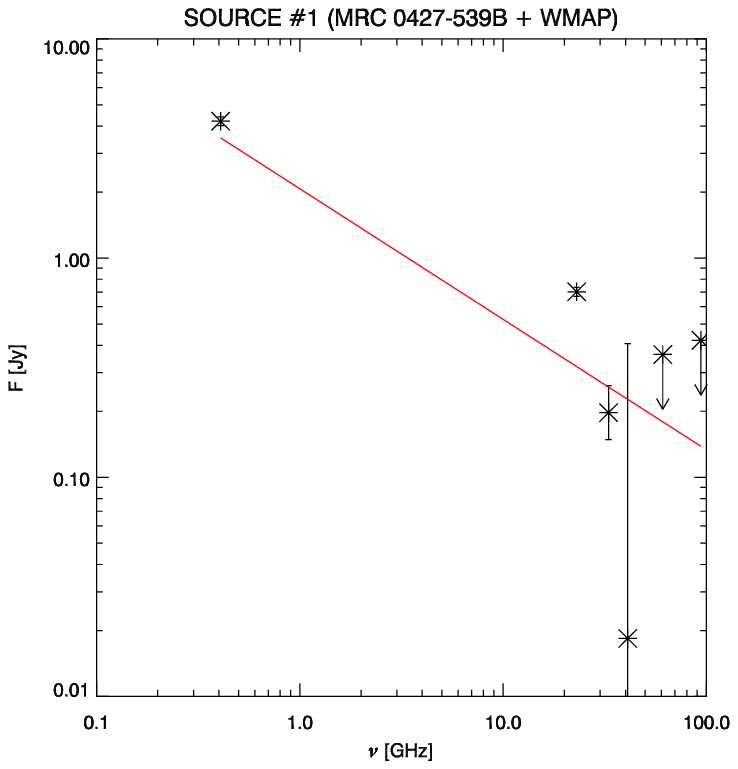}
	\includegraphics[height=8cm]{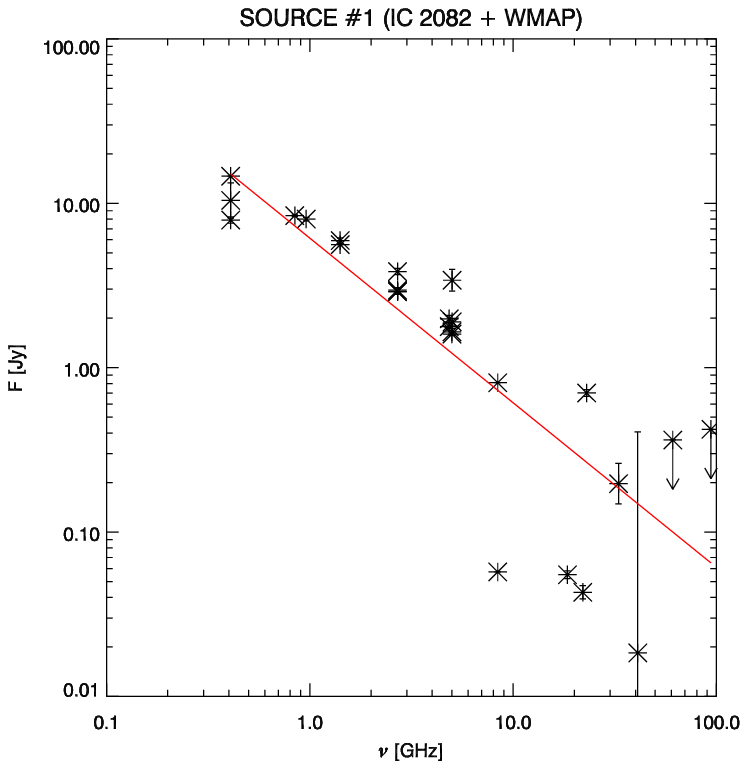}
\\
	\includegraphics[height=8cm]{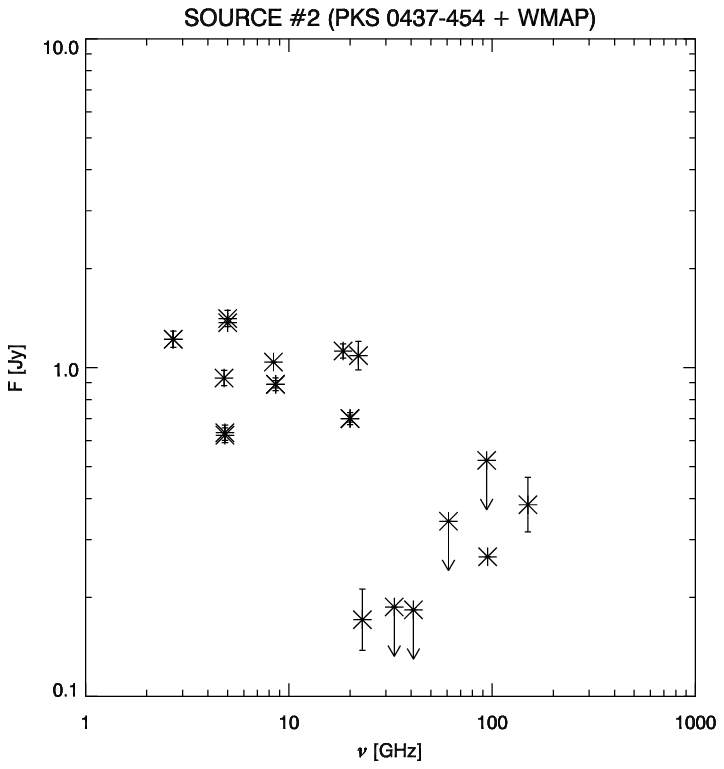}
	\includegraphics[height=8cm]{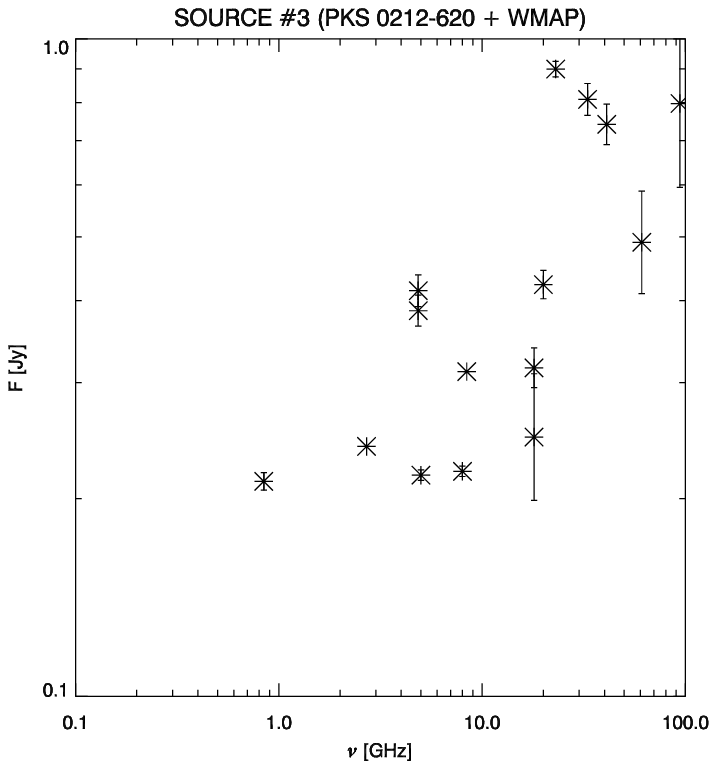}
\\
	\includegraphics[height=8cm]{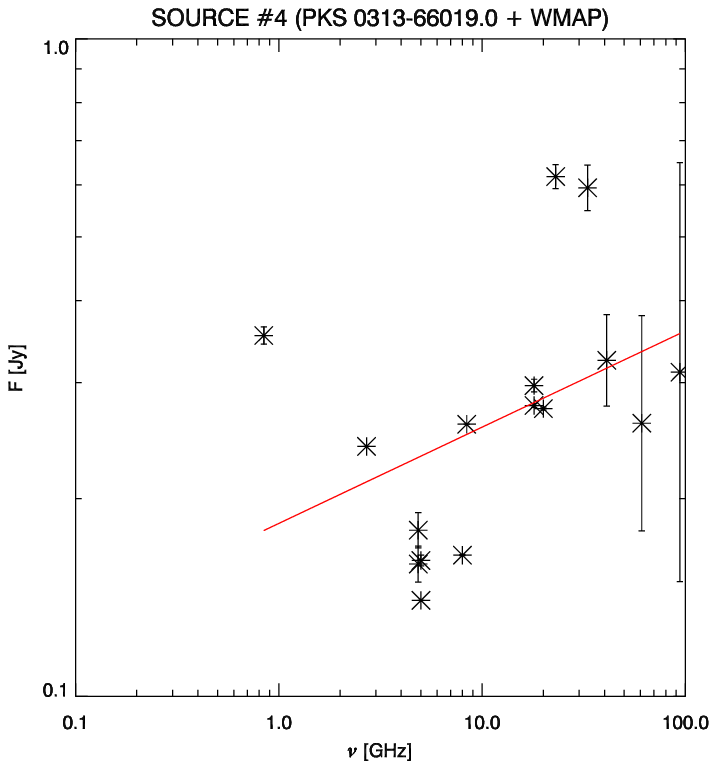}
	\includegraphics[height=8cm]{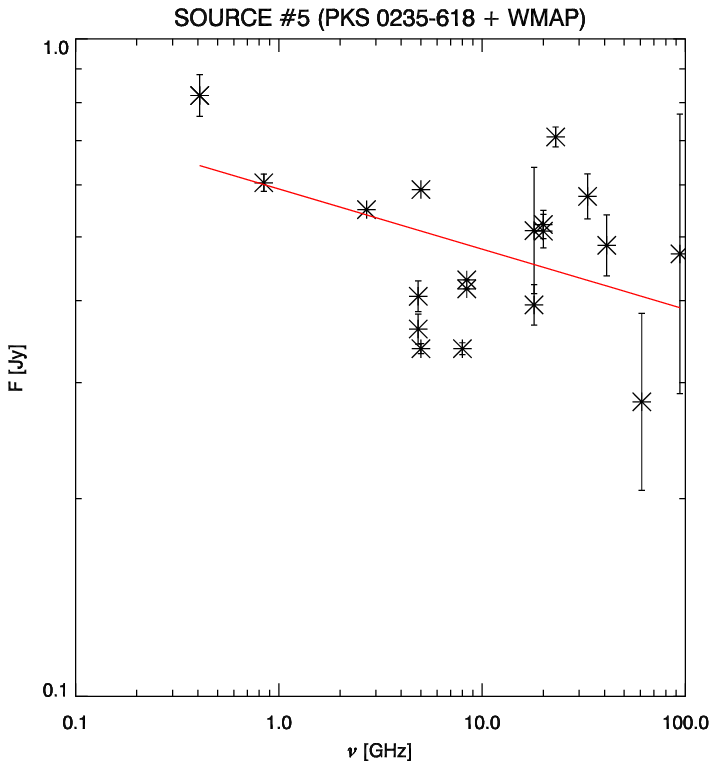}
\\
        \caption{Spectral Energy Distribution for the new sources using the derived WMAP data and NED data of the possible counterparts. The arrows represent upper limits corresponding to 3$\sigma$, where $\sigma$ was obtained from the pixel noise. The WMAP data is plotted with data from MRC 0427-539B and IC 2082 for {\it Source \# 1}, PKS 0437-454 for {\it Source \# 2}, PKS 0212-620 for {\it Source \# 3}, PKS 0313-66019.0 for {\it Source \# 4} and PKS 0235-618 for {\it Source \# 5}. For the {\it Sources \# 1, \# 4 and \# 5 }, it is also plotted the BCES(Y$|$X) ordinary least-squares regression lines.}
        \label{fig:flux_vs_freq_1_6} 
\end{figure*}

\begin{figure*} 
	\includegraphics[height=8cm]{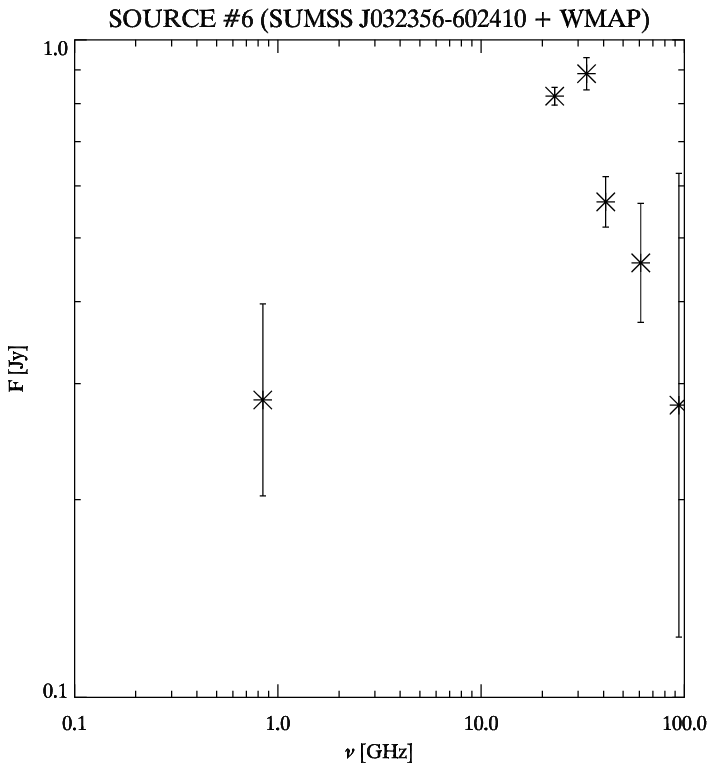}
	\includegraphics[height=8cm]{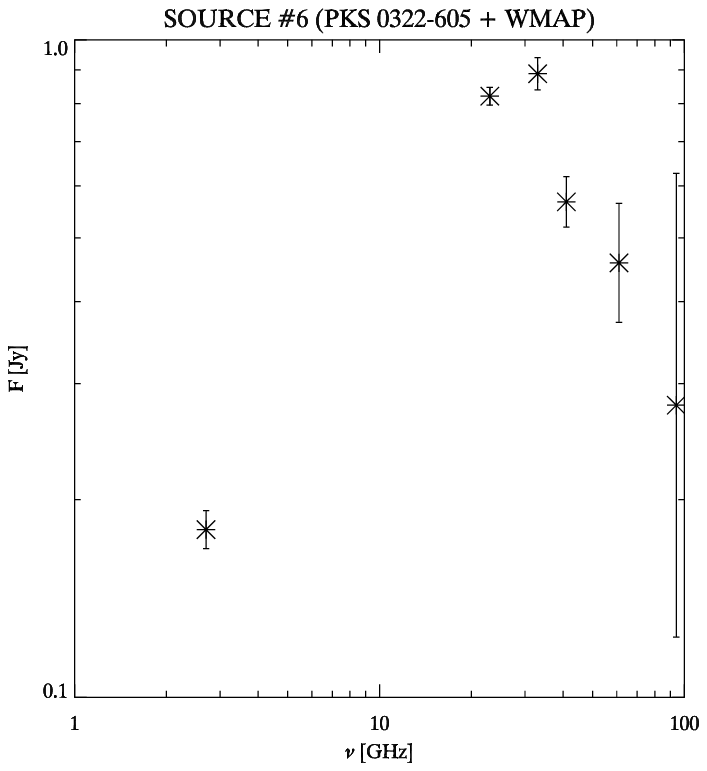}
\\
	\includegraphics[height=8cm]{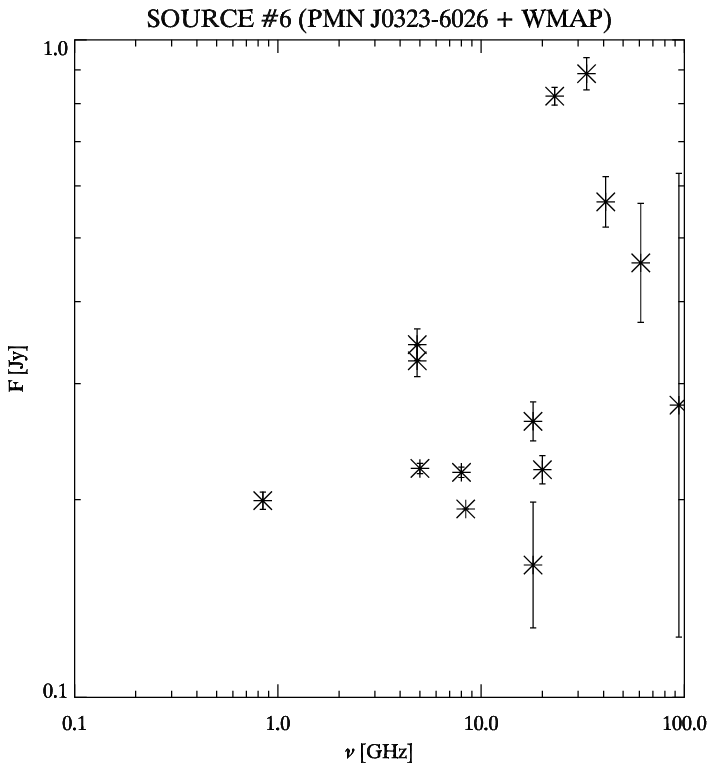}
	\includegraphics[height=8cm]{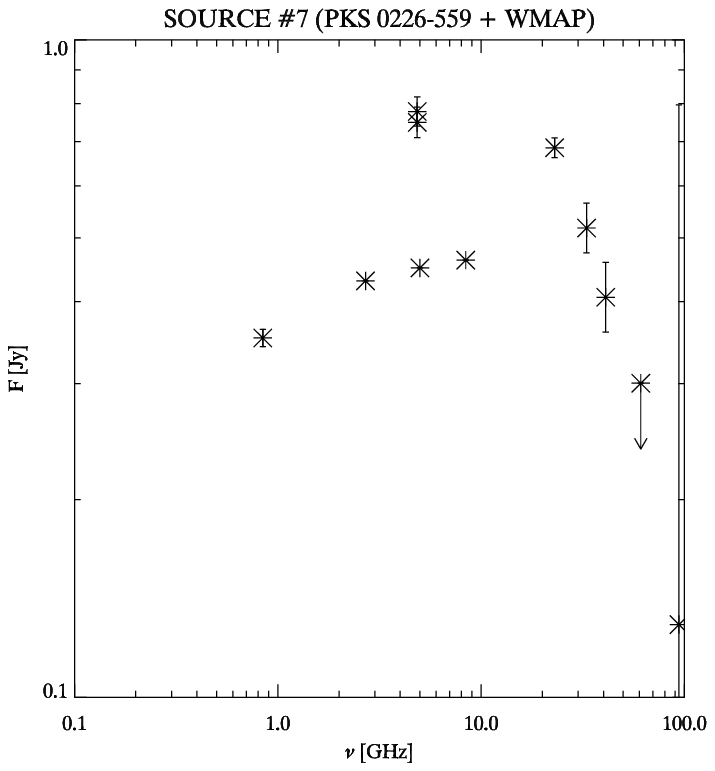}
	\caption{The same as Fig.~\ref{fig:flux_vs_freq_1_6}, with WMAP data plotted with data from SUMSS J032356-602410, PKS 0322-605 and PMN J0323-6026 for {\it Source \# 6} and PKS 0226-559 for {\it Source \# 7}.}
	\label{fig:flux_vs_freq_7_10}
\end{figure*}

We summarize in Tab.~\ref{tab:nedsrc} a list of possible counterparts. The Fig.~\ref{fig:flux_vs_freq_1_6} and Fig.~\ref{fig:flux_vs_freq_7_10} show the reconstructed radio Spectral Energy Distributions (SED) of the most likely counterparts with the WMAP data. The linear regression lines and respective spectral indices were obtained through the BCES(Y$|$X) ordinary least-squares method which takes into account measurement errors \citep{ab96}.
  
Here we discuss briefly the most likely identifications of the new sources, which we associate to the brightest radio sources within
our search radius. 
\begin{itemize}
\item
{\it Source \# 1} The strongest radio sources within our search radius are MRC 0427-539B of which only one flux value at 408 MHz is available and
IC 2082 that is a Galaxy pair (Gpair) with radio data from 408 MHz to 22 GHz.
We plot the corresponding radio SED of both sources assuming that the WMAP fluxes belong to them. Both spectra are plotted in
Fig.~\ref{fig:flux_vs_freq_1_6}.
In both cases we plot for reference also the best fit through the data which indicates a spectral behavior of an inverted-spectrum
with a spectral index of about -0.60$\pm$0.07 and -1.00$\pm$0.09, respectively.

\item
{\it Source \# 2} The most likely association is the radio source PKS 0437-454 which has data between 2.7 and 150 GHz.
Our reconstructed SED is reported in Fig.~\ref{fig:flux_vs_freq_1_6}. Here the distribution of fluxes is complex and may show 
a SED with different components. The uncertainties related to the identification and the source fluxes do not allow to
make any further investigation with the present data. 
 
\item
{\it Source \# 3} The closest radio source with the strongest fluxes is PKS 0212-620 which is a candidate QSO with radio data from 843 MHz to 20 GHz. The fluxes are shown in Fig.~\ref{fig:flux_vs_freq_1_6}. The source seems to be variable and differences in flux
values at the same frequency confirm this \citep[see][]{Sadler}. 

\item
{\it Source \# 4} The most likely association is PKS 0313-660 which has data from 843 MHz to 20 GHz and it is identified as a QSO.
 The SED built with WMAP data is reported in Fig.~\ref{fig:flux_vs_freq_1_6}
and the fit gives the spectral index of $\sim$ 0.15$\pm$0.01.

\item
{\it Source \# 5} The most probable identification is the QSO PKS 0235-618 that has radio data in the range 408 MHz to 20 GHz.
The SED built with WMAP data is reported in Fig.~\ref{fig:flux_vs_freq_1_6} resulting in a fit with a spectral index of $\sim$ 0.09$\pm$0.01.

\item
{\it Source \# 6} There are three sources with bright radio fluxes, namely SUMSS J032356-602410, PKS 0322-605 and
PMN J0323-6026. We report the radio SEDs in Fig.~\ref{fig:flux_vs_freq_7_10}. For the first and second possible counterpart,
only one flux value is available. The PMN J0323-6026 has data from 843 MHz to 20 GHz and is a variable QSO \citep[see][]{Sadler}.
\item
{\it Source \# 7} The closest and most likely association is PKS 0226-559 with radio data from 843 MHz to 8.4 GHz identified as
a flat spectrum (at frequencies smaller than 8GHz) radio QSO \citep{Healey}.
The reconstructed SED with the WMAP data is shown in Fig.~\ref{fig:flux_vs_freq_7_10}. At high frequencies the spectrum appears to
decrease with frequency and it is no longer flat.
\end{itemize}
Most of all the other radio possible counterparts for which we do not give any SED (see Tab.~\ref{tab:nedsrc}), have only one
radio detection mostly at 843 GHz of the order of mJy or do not have published radio data.

\section{Conclusions} \label{sec:conclusions}

In this paper we have presented a new technique, the {\it weighted matched filter} (WMF), to extract point-sources from astrophysical maps.
This method is quite simple to use and more robust in practical applications and it is optimal in extracting sources with a spectrum
different from a flat one (i.e. with a spectral index close to zero).
We have shown the reliability of this technique with some numerical simulations.

We have then applied the method to three Southern Hemisphere sky regions -- each with an area of 400 deg$^2$ --
of the seven-year WMAP temperature maps and compared the resulting sources with those
of the two seven-year WMAP point-sources catalogues.

We have found in these three regions seven additional sources not previously listed in WMAP catalogues and
discuss their most likely identification and spectral properties.

We plan to investigate and explore further the application of the WMF technique and the identification of the new sources in future experiments, namely with Planck observational data and with simulations aimed at reproducing the sky at the
ALMA frequencies and spatial resolution.
\\
\\
{\bf Acknowledgments} \\
E. P. Ramos was financially supported by a grant from Funda\c{c}\~ao para a Ci\^encia e a Tecnologia (POPH-QREN-SFRH/BD/45613/2008) and project PTDC/CTE-AST/64711/2006. E. P. Ramos and R. Vio would like to thank ESO for its hospitality and support through the DGDF funding programme.

\clearpage

{%
\begin{table*} 
\begin{center}
\newcommand{\mc}[3]{\multicolumn{#1}{#2}{#3}}
\begin{tabular} {c c c | c | c c | c |c | }\cline{4-8}
\textbf{} & \textbf{} & \textbf{} & \mc{5}{c|}{\textbf{Possible counterpart}}\\\hline
\mc{1}{|c}{\textbf{\#}} & \textbf{l ($^\circ$)} & \textbf{b ($^\circ$)} & \textbf{ID} & \textbf{l ($^\circ$)} & \mc{1}{c|}{\textbf{b ($^\circ$)}} & \mc{1}{c|}{\textbf{d (')}} & \mc{1}{c|}{\textbf{Type}}\\\hline
\mc{1}{|c}{\textbf{1}} & \textbf{262.44} & \textbf{-42.41} & MRC 0427-539B & 262.45 & \mc{1}{c|}{-42.41} & \mc{1}{c|}{0.39} & \mc{1}{c|}{RadioS}\\
\mc{1}{|c}{\textbf{}} &  &  & SUMSS J042900-534934 & 262.42 & \mc{1}{c|}{-42.37} & \mc{1}{c|}{2.51} & \mc{1}{c|}{RadioS}\\
\mc{1}{|c}{\textbf{}} &  &  & IC 2082 & 262.42 & \mc{1}{c|}{-42.35} & \mc{1}{c|}{3.49} & \mc{1}{c|}{Gpair}\\
\mc{1}{|c}{\textbf{}} &  &  & J042908-534940* & 262.42 & \mc{1}{c|}{-42.35} & \mc{1}{c|}{} & \mc{1}{c|}{}\\
\mc{1}{|c}{\textbf{}} &  &  & ABELL S0463 & 262.29 & \mc{1}{c|}{-42.36} & \mc{1}{c|}{7.21} & \mc{1}{c|}{GClstr}\\\hline
\mc{1}{|c}{\textbf{2}} & \textbf{250.75} & \textbf{-41.81} & APMCC 521 & 250.71 & \mc{1}{c|}{-41.78} & \mc{1}{c|}{2.73} & \mc{1}{c|}{GClstr}\\
\mc{1}{|c}{\textbf{}} & \textbf{} & \textbf{} & 1651** & 250.84 & \mc{1}{c|}{-41.84} & \mc{1}{c|}{} & \mc{1}{c|}{}\\
\mc{1}{|c}{\textbf{}} & \textbf{} & \textbf{} & PKS 0437-454 & 250.85 & \mc{1}{c|}{-41.76} & \mc{1}{c|}{5.47} & \mc{1}{c|}{VisS}\\
\mc{1}{|c}{\textbf{}} & \textbf{} & \textbf{} & J043900-452222* & 250.85 & \mc{1}{c|}{-41.76} & \mc{1}{c|}{} & \mc{1}{c|}{}\\\hline
\mc{1}{|c}{\textbf{3}} & \textbf{287.17} & \textbf{-52.71} & SUMSS J021241-615218 & 287.26 & \mc{1}{c|}{-52.71} & \mc{1}{c|}{3.23} & \mc{1}{c|}{RadioS}\\
\mc{1}{|c}{\textbf{}} & \textbf{} & \textbf{} & SUMSS J021309-615429 & 287.21 & \mc{1}{c|}{-52.65} & \mc{1}{c|}{3.43} & \mc{1}{c|}{RadioS}\\
\mc{1}{|c}{\textbf{}} & \textbf{} & \textbf{} & 832** & 286.98 & \mc{1}{c|}{-52.65} & \mc{1}{c|}{} & \mc{1}{c|}{}\\
\mc{1}{|c}{\textbf{}} & \textbf{} & \textbf{} & PKS 0212-620 & 286.96 & \mc{1}{c|}{-52.66} & \mc{1}{c|}{8.31} & \mc{1}{c|}{VisS***}\\
\mc{1}{|c}{\textbf{}} & \textbf{} & \textbf{} & J021416-614933* & 286.96 & \mc{1}{c|}{-52.66} & \mc{1}{c|}{} & \mc{1}{c|}{}\\\hline
\mc{1}{|c}{\textbf{4}} & \textbf{283.11} & \textbf{-45.29} & SUMSS J031459-655454 & 283.06 & \mc{1}{c|}{-45.28} & \mc{1}{c|}{2.50} & \mc{1}{c|}{RadioS}\\
\mc{1}{|c}{\textbf{}} & \textbf{} & \textbf{} & SUMSS J031406-654955 & 283.05 & \mc{1}{c|}{-45.40} & \mc{1}{c|}{7.03} & \mc{1}{c|}{RadioS}\\
\mc{1}{|c}{\textbf{}} & \textbf{} & \textbf{} & SUMSS J031431-660346 & 283.26 & \mc{1}{c|}{-45.21} & \mc{1}{c|}{7.63} & \mc{1}{c|}{RadioS}\\
\mc{1}{|c}{\textbf{}} & \textbf{} & \textbf{} & PKS 0313-66019.0 & 283.00 & \mc{1}{c|}{-45.40} & \mc{1}{c|}{7.94} & \mc{1}{c|}{QSO}\\
\mc{1}{|c}{\textbf{}} & \textbf{} & \textbf{} & J031422-654824* & 283.00 & \mc{1}{c|}{-45.40} & \mc{1}{c|}{} & \mc{1}{c|}{}\\
\mc{1}{|c}{\textbf{}} & \textbf{} & \textbf{} & SUMSS J031554-655309 & 282.94 & \mc{1}{c|}{-45.23} & \mc{1}{c|}{8.29} & \mc{1}{c|}{RadioS}\\
\mc{1}{|c}{\textbf{}} & \textbf{} & \textbf{} & SUMSS J031558-660153 & 283.08 & \mc{1}{c|}{-45.13} & \mc{1}{c|}{9.95} & \mc{1}{c|}{RadioS}\\\hline
\mc{1}{|c}{\textbf{5}} & \textbf{283.31} & \textbf{-51.26} & SUMSS J023639-613721 & 283.25 & \mc{1}{c|}{-51.30} & \mc{1}{c|}{3.19} & \mc{1}{c|}{RadioS}\\
\mc{1}{|c}{\textbf{}} & \textbf{} & \textbf{} & PKS 0235-618 & 283.20 & \mc{1}{c|}{-51.29} & \mc{1}{c|}{4.65} & \mc{1}{c|}{QSO}\\
\mc{1}{|c}{\textbf{}} & \textbf{} & \textbf{} & J023653-613615* & 283.20 & \mc{1}{c|}{-51.29} & \mc{1}{c|}{} & \mc{1}{c|}{}\\
\mc{1}{|c}{\textbf{}} & \textbf{} & \textbf{} & SUMSS J023738-614223 & 283.19 & \mc{1}{c|}{-51.16} & \mc{1}{c|}{7.40} & \mc{1}{c|}{RadioS}\\
\mc{1}{|c}{\textbf{}} & \textbf{} & \textbf{} & SUMSS J023706-613048 & 283.07 & \mc{1}{c|}{-51.35} & \mc{1}{c|}{10.31} & \mc{1}{c|}{RadioS}\\\hline
\mc{1}{|c}{\textbf{6}} & \textbf{275.89} & \textbf{-47.85} & SUMSS J032427-602924 & 275.92 & \mc{1}{c|}{-47.82} & \mc{1}{c|}{1.89} & \mc{1}{c|}{RadioS}\\
\mc{1}{|c}{\textbf{}} & \textbf{} & \textbf{} & SUMSS J032356-602410 & 275.86 & \mc{1}{c|}{-47.93} & \mc{1}{c|}{4.74} & \mc{1}{c|}{RadioS}\\
\mc{1}{|c}{\textbf{}} & \textbf{} & \textbf{} & PKS 0322-605 & 275.93 & \mc{1}{c|}{-47.97} & \mc{1}{c|}{7.64} & \mc{1}{c|}{RadioS}\\
\mc{1}{|c}{\textbf{}} & \textbf{} & \textbf{} & SUMSS J032518-603151 & 275.88 & \mc{1}{c|}{-47.71} & \mc{1}{c|}{8.05} & \mc{1}{c|}{RadioS}\\
\mc{1}{|c}{\textbf{}} & \textbf{} & \textbf{} & SUMSS J032331-602102 & 275.84 & \mc{1}{c|}{-48.00} & \mc{1}{c|}{9.17} & \mc{1}{c|}{RadioS}\\
\mc{1}{|c}{\textbf{}} & \textbf{} & \textbf{} & PMN J0323-6026 & 276.00 & \mc{1}{c|}{-47.98} & \mc{1}{c|}{9.27} & \mc{1}{c|}{QSO}\\
\mc{1}{|c}{\textbf{}} & \textbf{} & \textbf{} & J032308-602632* & 276.00 & \mc{1}{c|}{-47.98} & \mc{1}{c|}{} & \mc{1}{c|}{}\\
\mc{1}{|c}{\textbf{}} & \textbf{} & \textbf{} & SUMSS J032308-602511 & 275.97 & \mc{1}{c|}{-48.00} & \mc{1}{c|}{9.55} & \mc{1}{c|}{RadioS}\\
\mc{1}{|c}{\textbf{}} & \textbf{} & \textbf{} & SUMSS J032249-602546 & 276.01 & \mc{1}{c|}{-48.02} & \mc{1}{c|}{11.72} & \mc{1}{c|}{RadioS}\\\hline
\mc{1}{|c}{\textbf{7}} & \textbf{278.33} & \textbf{-56.45} & PKS 0226-559 & 278.23 & \mc{1}{c|}{-56.47} & \mc{1}{c|}{3.48} & \mc{1}{c|}{QSO}\\
\mc{1}{|c}{\textbf{}} & \textbf{} & \textbf{} & J022821-554603* & 278.23 & \mc{1}{c|}{-56.47} & \mc{1}{c|}{} & \mc{1}{c|}{}\\
\mc{1}{|c}{\textbf{}} & \textbf{} & \textbf{} & 912** & 278.43 & \mc{1}{c|}{-56.51} & \mc{1}{c|}{} & \mc{1}{c|}{}\\
\mc{1}{|c}{\textbf{}} & \textbf{} & \textbf{} & SUMSS J022736-555231 & 278.50 & \mc{1}{c|}{-56.45} & \mc{1}{c|}{5.70} & \mc{1}{c|}{RadioS}\\
\mc{1}{|c}{\textbf{}} & \textbf{} & \textbf{} & SUMSS J022827-555607 & 278.41 & \mc{1}{c|}{-56.33} & \mc{1}{c|}{7.50} & \mc{1}{c|}{RadioS}\\
\mc{1}{|c}{\textbf{}} & \textbf{} & \textbf{} & SUMSS J022847-555437 & 278.32 & \mc{1}{c|}{-56.32} & \mc{1}{c|}{7.75} & \mc{1}{c|}{RadioS}\\
\mc{1}{|c}{} & \textbf{} & \textbf{} & SUMSS J022805-553922 & 278.15 & \mc{1}{c|}{-56.57} & \mc{1}{c|}{9.70} & \mc{1}{c|}{RadioS}\\
\mc{1}{|c}{} & \textbf{} & \textbf{} & PMN J0228-5538 & 278.06 & \mc{1}{c|}{-56.57} & \mc{1}{c|}{11.76} & \mc{1}{c|}{RadioS}\\
\mc{1}{|c}{} & \textbf{} & \textbf{} & J022820-553725* & 278.06 & \mc{1}{c|}{-56.57} & \mc{1}{c|}{} & \mc{1}{c|}{}\\\hline
*AT20G Survey\\
**NEWPS 5yr 3-sigma Survey\\
 ***QSO candidate

\end{tabular}
\end{center}
\caption{New detected sources and possible counterparts. First column corresponds to the number of the new sources identified by arrows in Fig.~\ref{fig:region1} and Fig.~\ref{fig:region3} with the galactic coordinates (in degrees) in the second and third column. The fourth, fifth and sixth columns are the object name (ID) and galactic coordinates (in degrees) of the possible counterparts found in the NED database within a search radius of 12 arcminutes. The ID of the possible counterparts in the AT20G and NEWPS 5yr 3-sigma surveys is also present.The seventh column gives the distance (in arcminutes) between the WMAP coordinates and those of the possible counterparts and the last column gives the type of object.}
	\label{tab:nedsrc}
\end{table*}
}%

\clearpage

\clearpage

\clearpage

\appendix
\begin{flushright}
\section{The classic {\it{\textbf{matched filter}}} method to detect point-sources in CMB maps} \label{sec:efficient}
\end{flushright}

This appendix integrates the missing information of Sects.~2 and 3. We present here the classic {\it matched filter} method to detect point-sources
in the case of multi-frequency observations. The goal is to facilitate the comparison of such approach with that proposed in this work. Arguments will be developed starting from the one-dimensional signal and single frequency case. The two-dimensional signal and
multi-frequency case is developed in the second part.\\

\subsection{One-dimensional signal and single-frequency case}

In the case of the CMB observations, the following conditions are commonly assumed:
\begin{enumerate}
\item The point-sources have a known spatial profile $\ssb = a \gb$. The amplitude ``$a$'' is a scalar quantity 
different from source to source, whereas $\gb$ is a function which, due to the instrument beam, is identical for all the sources. Function $\gb$ is normalized in such a way that $\max{ \{ g[0], g[1], \ldots, g[N-1] \} } = 1$, where $N$ 
is length of the function; \\
\item The point-sources are embedded in a noise-background $\bf{n}$, i.e. the observed signal $\xb$ is given by $\xb = \ssb + \nb$. In
other words, noise is additive;\\
\item Noise $\nb$ is the realization of a stationary stochastic process with known covariance matrix
\begin{equation} \label{eq:C}
\Cb = {\rm E}[\nb \nb^T]. 
\end{equation}
Because of the Galactic contribution, especially at low Galactic latitudes, this hypothesis is not always satisfied but we assume that it
holds locally. This allows 
the computation of statistics such as the mean or the covariance matrix that, otherwise, should not be possible. 
Without loss of generality, it is assumed that ${\rm E}[\nb] = 0$.
\end{enumerate}   
Under these conditions, the detection problem consists in deciding whether $\xb$ is a pure
noise $\nb$ (hypothesis $H_0$) or it contains also the contribution of a source $\ssb$ 
(hypothesis $H_1$). In this way, the source detection problem is 
equivalent to a decision problem where two hypotheses hold:
\begin{equation} \label{eq:decision}
\left\{
\begin{array}{ll}
\Hc_0: & \quad \xb = \nb; \\
\Hc_1: & \quad \xb = \nb + \ssb.
\end{array}
\right.
\end{equation}
Under $\Hc_0$ the probability density function of $\xb$ is given by $p(\xb| \Hc_0)$ whereas under $\Hc_1$ by $p(\xb| \Hc_1)$.
At this point, it is necessary to fix the criterion to use for the detection, which depends on the particular case of interest. 
For example, one could decide that the non-detection or the misidentification of a bright source could be
more important than the detection of a fainter one, or viceversa. A very common and effective criterion is the Neyman-Pearson criterion
which consists in the maximization of the {\it probability of detection} $\PD$ 
under the constraint that the {\it probability of false alarm} $\PFA$ (i.e., the probability of a false detection) does not exceed a fixed value
$\alpha$. The Neyman-Pearson theorem \citep[e.g., see ][]{kay98} is a powerful tool that allows to design
a decision process that pursues this aim:
{\it to maximize $\PD$ for a given $\PFA=\alpha$, decide $\Hc_1$ 
if the {\it likelihood ratio} (LR) 
\begin{equation} \label{eq:ratio}
L(\xb) = \frac{p(\xb| \Hc_1)}{p(\xb| \Hc_0)} > \gamma,
\end{equation}
where the threshold $\gamma$ is found from
\begin{equation} \label{eq:p1}
\PFA = \int_{\{\xb: L(\xb) > \gamma\}} p(\xb| \Hc_0) d\xb = \alpha.
\end{equation}
}
The test of the ratio~\eqref{eq:ratio} is called the {\it likelihood ratio test} (LRT). 

\begin{figure*}
        \resizebox{\hsize}{!}{\includegraphics{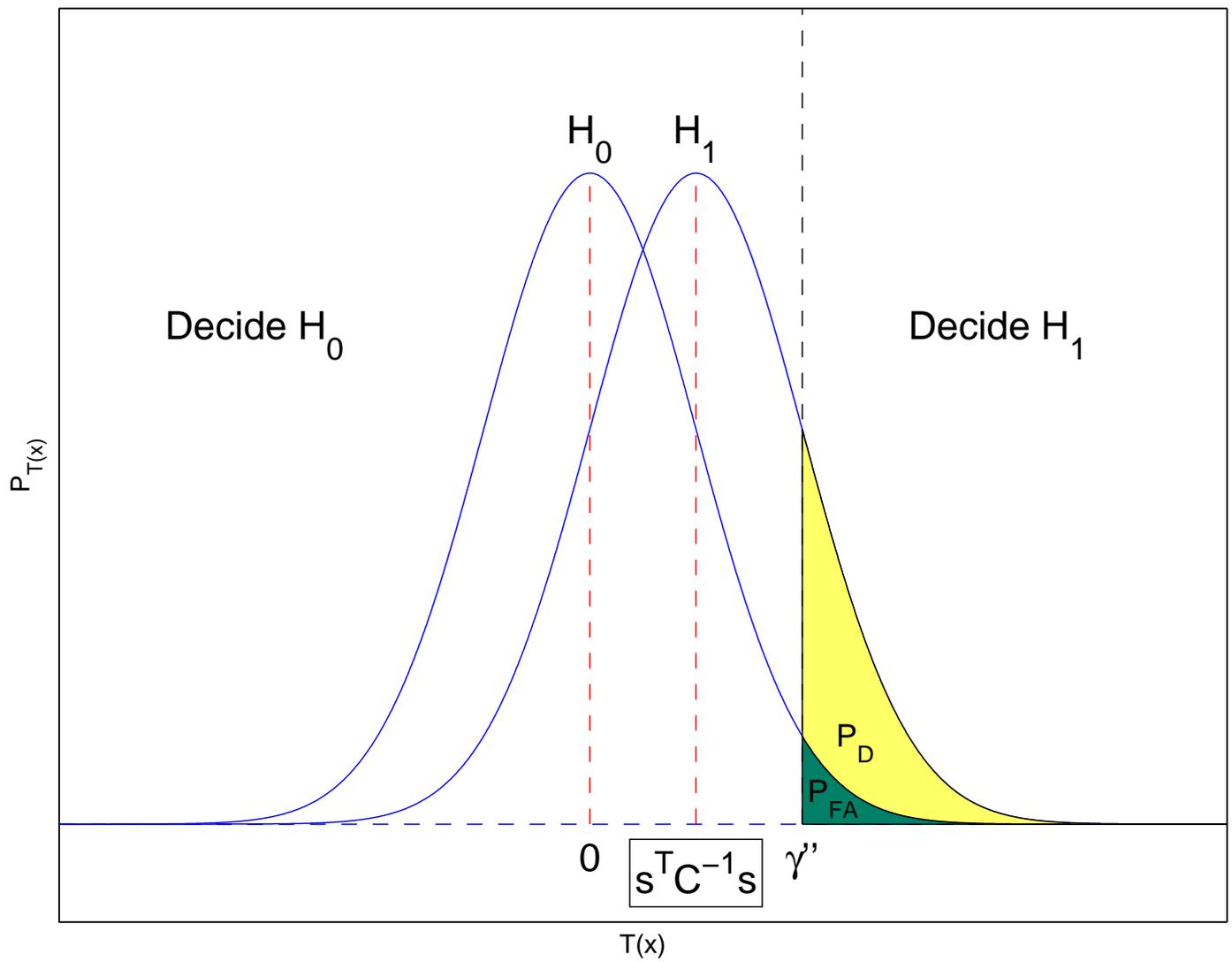}}
        \caption{Probability density function of the statistic $T(\xb)$ under the hypothesis $\Hc_0$ (noise-only hypothesis) and $\Hc_1$ (signal-present hypothesis). The detection-threshold is given by $\gamma^{\prime\prime}$. The {\it probability of false alarm} ($\PFA$) 
and the {\it probability of detection} ($\PD$) are shown in green and yellow colors, respectively.} 
        \label{fig_tx}
\end{figure*}

An important example of application of LRT is the case of a Gaussian noise $\nb$ with correlation function
$\Cb$. Actually, in CMB experiments this condition is satisfied only for observations at high Galactic latitudes where the CMB emission
and the instrumental noise are the dominant contributions. At lower latitudes, it is often assumed 
to hold locally. For example, the contribution to $\xb$ of components that in small sky patches show linear spatial 
trends are often approximated with stationary Gaussian processes with a steep spectrum (e.g. $1/f$ noises).
This is usually assumed, even in cases of unrealistic Gaussianity condition,
since it allows an analytical treatment of the problem 
of interest and the results can be used as a benchmark in the analysis of more complex scenarios. When $\nb$ is Gaussian,
\begin{align}
p(\xb | \Hc_0) & = \Delta
\exp\left[ -\frac{1}{2} \xb^T \Cb^{-1} \xb \right]; \label {eq:t1} \\
p(\xb | \Hc_1) & = \Delta  
\exp\left[ -\frac{1}{2} (\xb - \ssb)^T \Cb^{-1} (\xb -\ssb) \right], \label{eq:t2}
\end{align}
with
\begin{equation}
\Delta = \frac{1}{(2 \pi)^{\frac{N}{2}} {\rm det}^{\frac{1}{2}}(\Cb)}.
\end{equation}
The LRT is given by
\begin{equation}
l(\xb) = \ln[L(\xb)] = \xb^T \Cb^{-1} \ssb - \frac{1}{2} \ssb^T \Cb^{-1} \ssb > \gamma'. 
\end{equation}
Hence, it results that
$\Hc_1$ has to be chosen when the statistic $T(\xb)$ (called {\it NP detector}) is
\begin{equation} \label{eq:test}
T(\xb) = \xb^T \Cb^{-1} \ssb > \gamma'',
\end{equation}
with $\gamma''$ such that
\begin{equation} \label{eq:pfa}
\PFA = Q \left( \frac{\gamma''}{\left[ \ssb^T \Cb^{-1} \ssb \right]^{1/2}} \right) = \alpha,
\end{equation}
i.e.,
\begin{equation} \label{eq:gammap}
\gamma'' = Q^{-1}(\PFA) \sqrt{ \ssb^T \Cb^{-1} \ssb }.
\end{equation}
Here, $Q(x) $ is the {\it complementary cumulative distribution function
\begin{equation}
Q(x) = \int_x^{\infty} \frac{1}{\sqrt{2 \pi}} \left( \exp{- \frac{1}{2} t^2} \right) dt,
\end{equation}
and $Q^{-1}$
the corresponding inverse function}. Equation~\eqref{eq:pfa} is due to the fact that $T(\xb)$ is a Gaussian random variable with 
variance $\ssb^T \Cb^{-1} \ssb$ and expected values equal to zero under $\Hc_0$ and $\ssb^T \Cb^{-1} \ssb$ under $\Hc_1$
(see Fig.~\ref{fig_tx}).
For the same reason, 
the $\PD$ is given by
\begin{equation} \label{eq:pd}
\PD = Q \left( Q^{-1} \left( \PFA \right) - \sqrt{ \ssb^T \Cb^{-1} \ssb} \right).
\end{equation}
Equation~\eqref{eq:test} can be written in the form
\begin{equation} \label{eq:test1}
T(\xb) = \xb^T \ub > \gamma'',
\end{equation}
with 
\begin{equation} \label{eq:mf}
\ub = \Cb^{-1} \ssb.
\end{equation}
From this equation $\ub$ can be thought as a linear filter of signal $\xb$. It is called {\it matched filter} (MF).

\subsubsection{Some comments on the use of the {\it matched filter} in practical applications} \label{sec:comments}

There are some important points to stress about the MF when used in practical applications, such as:
\begin{itemize}
\item $T(\xb)$ is a {\it sufficient statistic} \citep{kay98}.
Loosely speaking, this means that $T(\xb)$ is able to summarize all the relevant information in the data
concerning the decision~(\ref{eq:decision}). No other statistic can perform better; \\

\item If the amplitude ``$a$'' of the source is unknown,
then Eq.~(\ref{eq:test}) can be rewritten in the form
\begin{equation} \label{eq:test2}
T(\xb) = \xb^T \Cb^{-1} \gb > \gamma''',
\end{equation}
with $\gamma''' = \gamma'' / a = Q^{-1}(\PFA) \sqrt{ \gb^T \Cb^{-1} \gb }$. In other words, a statistic is  
obtained independent of ``$a$''. As a consequence of the Neyman-Person theorem,
in the case of unknown amplitude of the source, $T(\xb)$ still maximizes $\PD$
for a fixed $\PFA$. The only consequence is that $\PD$ cannot be evaluated in advance. In principle this
can be done a posteriori by using the maximum likelihood estimate of the amplitude, 
$\widehat{a} = \xb^T \Cb^{-1} \gb / \gb^T \Cb^{-1} \gb$. However, this is of little interest,  
since in real experiments
one is typically interested in the detection of sources which have amplitudes characterized by a probability density function 
$p(a)$. In this case, once $\PFA$ is fixed to a
value $\alpha$ and changing ``$a$'' across the domain of $p(a)$, the quantity $1-\PD$, with $\PD$ given by 
Eq.~(\ref{eq:pd}) and $\ssb = a \gb$, provides an estimate of the fraction of undetected sources as function of their amplitude; \\

\item If $\shb = \Hb \ssb$ and $\xhb = \Hb \xb$, then
\begin{align}
T(\Hb \xb) & = \xhb^T \Chb^{-1} \shb \nonumber  \\
& = \xb^T \Hb^T \Hb^{-T} \Cb^{-1} \Hb^{-1} \Hb \ssb = T(\xb),
\end{align}
with $\Hb$ any invertible linear operator (matrix).
A useful consequence of this property is that if signal $\xb$ is convolved with a function (e.g., the beam of an instrument), 
this operation does not modify the optimality of MF. In this case, the operator $\Hb$ transforms the covariance
matrix $\Cb$ into $\Hb \Cb \Hb^T$. This fact is useful in situations where more signals are available
that are obtained with different point spread functions.
\end{itemize}

\subsection{One-dimensional signal and multiple-frequency case} \label{sec:multiple}

In the context of CMB observations, the complexity increases since there are $M$ signals $\xb_k = \ssb_k + \nb_k$,
$\ssb_k = a_k \gb_k$, $k = 1, 2, \ldots, M$,
coming from the same sky area that are taken at different observing frequencies. Here, $a_k$ is the amplitude of the source at the
$k$th observing frequency, whereas $\gb_k$ is the corresponding spatial profile. For ease of notation, all the signals are assumed to have
the same length $N$. In general, the amplitudes $\{ a_k \} $ as well as the
profiles $\{ \gb_k \}$ are different for different $k$. However, if one sets
\begin{align}
\xb & = [\xb_1^T, \xb_2^T, \ldots, \xb_M^T]^T, \label{eq:mxb} \\
\ssb & = [\ssb_1^T, \ssb_2^T, \ldots, \ssb_M^T]^T, \label{eq:msb} \\
\nb & = [\nb_1^T, \nb_2^T, \ldots, \nb_M^T]^T, \label{eq:mnb}
\end{align}
it is possible to obtain a problem that is
formally identical to that treated in the previous section. Hence, the MF is still given by Eqs.~\eqref{eq:test1}-\eqref{eq:mf} and
is named {\it multi-frequency matched filter} (MMF). The only difference with the classic MF is that now $\Cb$ is a $(N M) \times (N M)$ 
{\it block matrix with Toeplitz blocks} (BTB):
\begin{equation} \label{eq:covariance}
\Cb = 
\left( \begin{array}{ccc}
\Cb_{11} & \ldots & \Cb_{1M} \\
\vdots & \ddots & \vdots \\
\Cb_{M1} & \ldots & \Cb_{MM} \\
\end{array} \right),
\end{equation}
i.e., each of the $\Cb_{ij}$ blocks is constituted by a $N \times N$ Toeplitz matrix. In particular, 
$\Cb_{ii} = {\rm E}[\nb_i \nb_i^T]$ provides the autocovariance matrix of 
the $i$th noise, whereas $\Cb_{ij} = {\rm E}[\nb_i \nb_j^T]$, $i \neq j$, the cross-covariance matrix between the $i$th and the $j$th ones.

In spite of these similarities, when $M > 1$, there are additional difficulties: $T(\xb)$ cannot be written in a form equivalent to
Eq.~\eqref{eq:test2}. In the case of unknown amplitudes $\{ a_k \}$ $T(\xb)$
cannot be computed. In other words, if the spectral characteristics of the radiation emitted by a source are not fixed, then the MMF is not 
applicable. 

\subsection{Extension to two-dimensional signals}

The extension of MF to the two-dimensional signals $\Xmatb$ and $\Smatb$ is conceptually trivial. In the case of one-dimensional case, setting
\footnote{${\rm VEC}[\Fmatb]$ is the operator that transforms a matrix $\Fmatb$ into a column array by stacking its columns one underneath
the other.}
\begin{align}
\ssb & = {\rm VEC}[\Smatb]; \label{eq:stack1} \\
\xb & = {\rm VEC}[\Xmatb]; \label{eq:stack2} \\
\nb & = {\rm VEC}[\Nmatb], \label{eq:stack3}
\end{align}
formally the problem results in the same as given by Eq.~\eqref{eq:decision}.
In the case of multi-frequency observations, the situation is more complex
since MF has to be applied to $M$ signals at the same time. However, with the notation
\begin{align}
\ssb & = {\rm VEC}\left[ {\rm VEC}[ \Smatb_1 ], {\rm VEC}[ \Smatb_2 ], \ldots, {\rm VEC}[ \Smatb_M ] \right]; \label{eq:stack12} \\
\xb & = {\rm VEC}\left[ {\rm VEC}[ \Xmatb_1 ], {\rm VEC}[ \Xmatb_2 ], \ldots, {\rm VEC}[ \Xmatb_M ] \right]; \label{eq:stack22} \\
\nb & = {\rm VEC}\left[ {\rm VEC}[ \Nmatb_1 ], {\rm VEC}[ \Nmatb_2 ], \ldots, {\rm VEC}[ \Nmatb_M ] \right], \label{eq:stack32}
\end{align}
it is also possible to obtain a problem that is
formally identical to that given by Eq.~\eqref{eq:decision}. The only difference is that now in Eq.~\eqref{eq:covariance}, 
$\Cb$  is a $(M N_p) \times (M N_p)$ 
block matrix. If the signals $\{ \Smatb_i \}$ are two-dimensional $N_r \times N_c$ maps, then  
each of the $\Cb_{ij}$ blocks is constituted by a $(N_r N_c) \times (N_r N_c)$  {\it block Toeplitz with Toeplitz blocks} (BTTB) matrix. In particular, 
$\Cb_{ii}$ provides the autocovariance matrix of
the $i$th map, whereas $\Cb_{ij}$, $i \neq j$, the cross-covariance matrix between the $i$th and the $j$th maps. 

Especially for the multi-frequency case, the implementation of the MF is not trivial. Even for moderate size signals, the matrix $\Cb$
becomes rapidly huge. As a consequence, it is necessary to implement numerical methods which are able to exploit the specific 
structure of $\Cb$. Typically, they are based on Fourier approaches. This subject, however, is beyond the aim of the present work. 

\end{document}